\documentclass[%
 reprint,superscriptaddress,
 twocolumn,
 amsmath,amssymb,
 aps,prx,
]{revtex4-2}

\usepackage[utf8]{inputenc}
\usepackage[hidelinks,hypertexnames=false, colorlinks=true,allcolors=blue]{hyperref}
\usepackage[all]{hypcap}

\usepackage{graphicx,float}
\usepackage{amsmath}
\usepackage{amssymb}
\usepackage{bm}
\usepackage{amsfonts}
\usepackage{appendix}

\usepackage{dsfont}
\usepackage{txfonts}

\usepackage[dvipsnames]{xcolor}
\usepackage{upgreek}
\usepackage{physics}
\usepackage[normalem]{ulem}

\usepackage[utf8]{inputenc} 
\usepackage[T1]{fontenc}

\newcommand{\e}{\textrm{e}}
\renewcommand{\i}{\textrm{i}}

\newcommand{\Jeff}{J_{\mathrm{eff}}}
\newcommand{\zcm}{z_{\mathrm{cm}}}
\newcommand{\pcm}{p_{\mathrm{cm}}}
\newcommand{\Er}{ E_\text{r} }
\newcommand{\Nex}{N_\mathrm{ex}}
\newcommand{\Ntot}{N_\mathrm{tot}}
\newcommand{\kL}{k_\mathrm{L}}
\newcommand{\dL}{d_\mathrm{L}}
\newcommand{\TD}{T_\mathrm{D}}
\newcommand{\FD}{F_\mathrm{D}}
\newcommand{\Kc}{K_\mathrm{c}}
\newcommand{\omgd}{\omega_\mathrm{D}}

\begin{document}

\title{Suppressing Parametric Instabilities in Driven Bosonic Lattices through Multi-tone Control}

\author{Robbie Cruickshank}\thanks{These authors contributed equally to this work.}
\affiliation{Department of Physics and SUPA, University of Strathclyde, Glasgow G4 0NG, United Kingdom}
\author{Samuel Lellouch}\thanks{These authors contributed equally to this work.}
\affiliation{School of Physics and Astronomy, University of Birmingham, Birmingham B15 2TT, United Kingdom}
\affiliation{
School of Engineering, University of Birmingham, Birmingham B15 2TT, United Kingdom}
\author{Marin Bukov}\thanks{These authors contributed equally to this work.}
\affiliation{Max Planck Institute for the Physics of Complex Systems, Nöthnitzer Str. 38, 01187 Dresden, Germany}
\author{Eugene Demler}
\affiliation{Institute for Theoretical Physics, ETH Zürich, 8093 Zürich, Switzerland}
\author{Nathan Goldman}
\email{nathan.goldman@lkb.ens.fr}
\affiliation{Laboratoire Kastler Brossel, Coll\`ege de France, CNRS, ENS-Universit\'e PSL,
Sorbonne Universit\'e, 11 Place Marcelin Berthelot, 75005 Paris, France
}
\affiliation{International Solvay Institutes, 1050 Brussels, Belgium}
\affiliation{Center for Nonlinear Phenomena and Complex Systems, Universit\'e Libre de Bruxelles, CP 231, Campus Plaine, B-1050 Brussels, Belgium}
\author{Elmar Haller}
\email{elmar.haller@strath.ac.uk}
\affiliation{Department of Physics and SUPA, University of Strathclyde, Glasgow G4 0NG, United Kingdom}
\date{\today}

\begin{abstract}
Periodically driven quantum systems offer remarkable flexibility in tailoring effective Hamiltonians and synthetic band structures. However, such driving also induces heating and dynamical instabilities that limit the coherence and lifetime of many-body states. Here, we demonstrate that these instabilities can be suppressed by employing multi-tone driving schemes. Using a Bose-Einstein condensate of cesium atoms in an optical lattice, we experimentally explore two approaches: pulsed driving composed of odd harmonics and two-tone driving with tunable amplitude and relative phase. We show that both methods allow independent control of the effective tunneling amplitude and Peierls phase factor, while significantly reducing phonon excitation and the resulting rapid decay of the condensate. Numerical simulations and theoretical modeling based on Bogoliubov-de Gennes equations confirm the suppression of unstable modes under optimized driving conditions. Our results establish multifrequency drives as powerful tools for stabilizing driven many-body systems and pave the way toward robust Floquet engineering with interactions.
\end{abstract}

\maketitle

\section{Introduction\label{sec:Introduction}}
Periodic drives constitute an essential toolbox in contemporary quantum simulation, enabling the engineering of synthetic Hamiltonians with tailored properties~\cite{goldman2014,eckardt2017,schindler2024counterdiabatic}. 
For neutral-atom simulators, such as ultracold atomic gases in optical lattices or analog Rydberg array platforms, this \textit{Floquet engineering} approach has led to breakthroughs, such as the realization of artificial gauge fields, tunable tunneling amplitudes, and topological band structures~\cite{goldman2014b, goldman2016topological, aidelsburger2018artificial, cooper2019topological}. A key imminent milestone for advancing the capabilities of quantum simulation concerns the reliable realization of strongly-correlated states of quantum matter~\cite{daley2022practical}, such as doped quantum magnets in the Fermi-Hubbard model~\cite{BOHRDT2021168651,chalopin2024probing,xu2025neutral} and topologically-ordered states in quantum-Hall and spin-liquid quantum-engineered settings~\cite{leonard2023realization,palm2024growing,impertro2025strongly,kalinowski2023non-abelian,sun2023engineering,chen2024realization,evered2025probing,will2025probing}. However, incorporating interactions in Floquet systems introduces significant challenges by opening up additional channels for energy absorption, mode coupling, and collective instabilities.

Experimentally, periodic drives used to engineer exotic interacting states usually cause rapid heating, exciting the system before the dynamics can fully manifest the controlled effective Hamiltonian~\cite{miyake2013,weinberg2015multiphoton,reitter2017, messer2018,rubioabadal2020,dicarli2023, cruickshank2024}.   
In particular, for systems with bosonic excitations, the presence of weak interactions is sufficient to induce parametric instabilities~\cite{ gluck2002wannier, bukov2012,lellouch2017,boulier2019, wintersperger2020, shavit2025parametric} seeded by quantum and thermal fluctuations or technical noise in the laser phase and amplitude~\cite{tozzo2005}. The exponentially growing population of parametrically amplified modes results in rapid condensate depletion, momentum distribution broadening, and loss of quantum coherence~\cite{lellouch2017}; it accelerates heating via nonlinear scattering beyond the low-energy Bogoliubov description. 
At long times, the growth saturates, giving rise to a metastable state due to nonlinear interactions, whose momentum distribution patterns have been observed experimentally~\cite{dupont2023emergence, liebster2025observation}. Crucially, from the perspective of Floquet-engineering, redistributed absorbed energy is tantamount to irreversible heating to a high-entropy state~\cite{bukov2015universal}.

Theoretically, parametric resonances lead to the early breakdown of inverse-frequency expansions, and preclude exponentially long-lived prethermal plateaus~\cite{bukov2016heating, abanin2017rigorous, kuwahara2016floquet} from forming, thereby shrinking or entirely eliminating the playground for Floquet engineering. Hence, their analysis poses formidable challenges within the Floquet-engineering formalism, and highlights the important role of experiments to advance theory.
For all these reasons, finding protocols to eliminate or suppress parametric resonances is quintessential for the simulation of interacting models with bosonic excitations.
    
One way to do this is to notice that the Floquet engineering toolbox is not limited to conceptually minimal drives. The Fourier space of the real-time protocol admits a vast and largely unexplored parameter space, with many possible combinations of frequency and amplitude schedules. 
Only recently have studies started to investigate drive engineering as a possible method to provide additional experimental control. 
Beyond single-tone sinusoidal driving, recent studies have explored alternative schemes, including square-wave forcing~\cite{eckardt2009}, analyzed more naturally using Walsh functions~\cite{votto2024universal,walkling2025walsh}, polychromatic driving~\cite{verdeny2015, grossert2016,murakami2023,impens2024}, and interaction modulation~\cite{zhao2019}. Specifically, two-tone drives have been proposed to introduce synthetic dimensions to realize nontrivial topological band structures~\cite{martin2017}, heating suppression~\cite{banerjee2024}, for band-gap control~\cite{sandholzer2022,wang2023,chen2025}, and to engineer inherently-nonequilibrium states of matter such as time crystals~\cite{beatrez2023critical,moon2025experimental}.
Here we raise the question of how to design periodic drives which, on the one hand, possess Floquet engineering capabilities, and on the other, suppress the detrimental parametric instabilities for interacting bosons in optical lattices. 

Specifically, we focus on the role of interactions for multi-frequency drives in the lowest lattice band of a one-dimensional lattice. We demonstrate experimentally that the parametric growth of excitation modes can be suppressed while preserving the advantages of periodic driving through tailored drive engineering. We investigate two alternative protocols that incorporate multiple driving tones: (i) pulsed driving, where the force is applied in short, discrete bursts, and (ii)~two-tone driving in which both the strength and relative phase between the components are varied. We show that these approaches effectively suppress the growth of unstable modes while maintaining the full flexibility of Floquet engineering to control the effective band structure of the lattice. 

For both driving protocols, we characterize the condensates' stability across a range of driving parameters and identify robust regimes where the parametric growth of phonon excitations is substantially suppressed. The reduction is significant, far beyond what is achievable with sinusoidal driving, while maintaining the same effective tunneling and band structure. Our measurements show excellent agreement with analytical predictions and numerical simulations, providing theoretical insight into the underlying mechanisms responsible for the enhanced stability.
This work highlights the promising potential of multi-frequency periodic drives to alleviate the difficulties associated with achieving a stable simulation of strongly interacting systems with bosonic excitations.  

This article is structured into four main sections. Following this introduction, we present the theoretical description and experimental realization of our system in Sect.~\ref{sec:Exp_Methods}, including the driving mechanism and the growth of phonon modes. Section \ref{sec:Pulsed_Drive} introduces the pulsed driving scheme and demonstrates its advantages, while Sect.~\ref{sec:TwoFreq_Drive} focuses on the implementation and the impact of the two-tone drive on phonon modes. 
In Sect.~\ref{sec:PeierlsPhase}, we show that both driving mechanisms also enable control over the complex phases of the tunneling matrix elements, resulting in gauge potentials that couple to the momentum of the atoms. Finally, Sect.~\ref{sec:Conclusion} offers a comparative summary of the results and discusses their implications.

\section{Driven lattice system \label{sec:Exp_Methods}}

\begin{figure}[t] \centering
\includegraphics[width=1\columnwidth]{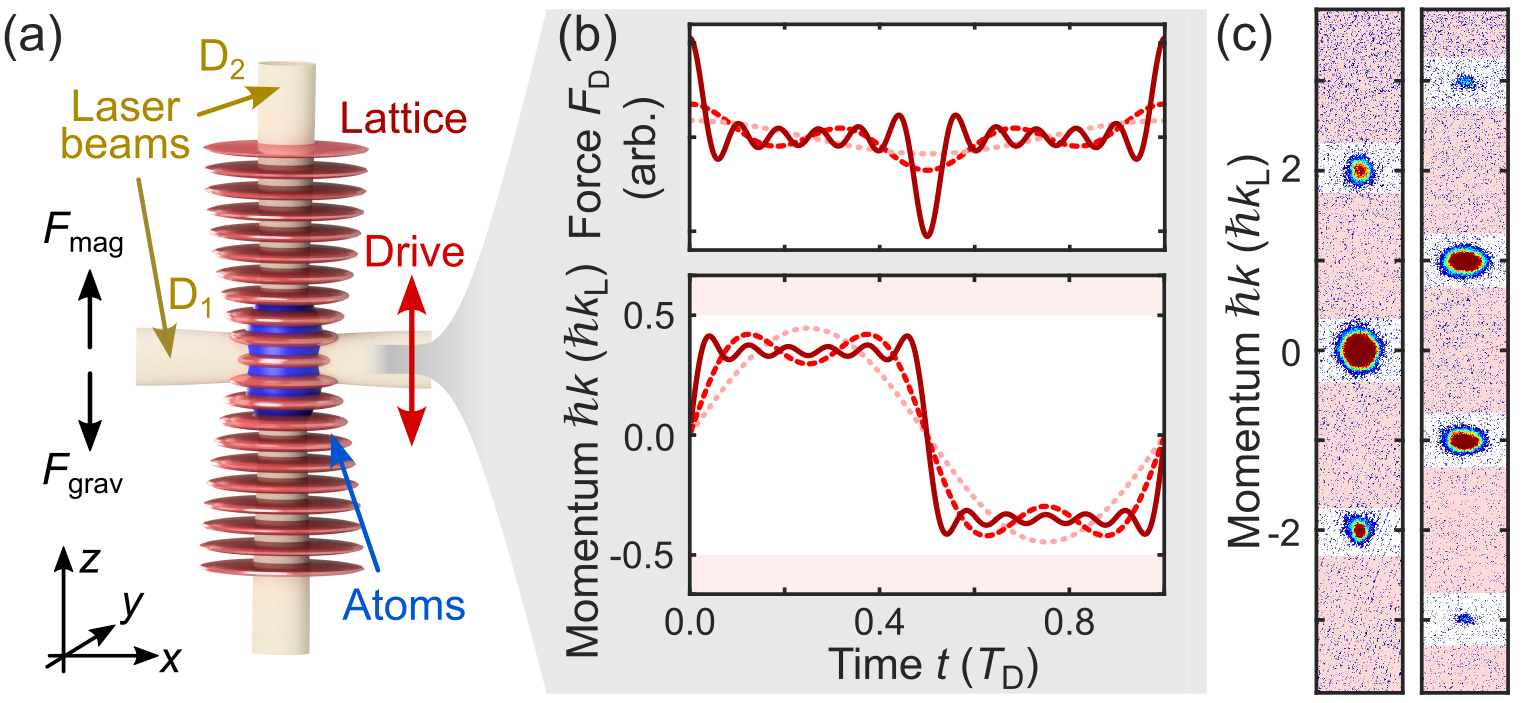}
\caption{Experimental methods and setup. (a) Schematic of the experimental apparatus, showing guiding laser beams $D_1$ and $D_2$, and the optical lattice. The driving force is realized in the lattice frame by periodically shifting the lattice sites. (b) Calculated driving force ($\FD$, top) and micromotion in momentum space ($\hbar k[t]$, bottom)  over one driving period $\TD$, shown for pulsed driving with $M=1, 3, 11$ harmonics (dotted, dashed, and solid lines). Red-shaded areas highlight regions of the Brillouin zone with $k\geq0.5\kL$. (c) Absorption images in the laboratory frame showing the atomic momentum distribution after time-of-flight expansion for initial momentum $\hbar k_0=0$ (left) and $\hbar k_0=\hbar \kL$ (right) at $t=0$. Atoms found in the red-shaded momentum regions are counted to determine the excitation fraction $\Nex/\Ntot$. Radial excitations remain weak over the applied modulation time and are neglected in the analysis. \label{Fig:Setup} }
\end{figure}

Our system consists of a Bose-Einstein condensate of approximately $N=80,000$ cesium atoms confined in a one-dimensional optical lattice potential [Fig.~\ref{Fig:Setup}(a)]. The lattice is created by two counter-propagating laser beams at a wavelength of $\lambda = 1064\,$nm, yielding a spacing of $\dL = \lambda/2$ and a corresponding lattice momentum $\kL = \pi/\dL$. Within the mean-field approximation, the system is well described by the wavefunction $\psi(\mathbf{r},t)$ and the Gross-Pitaevskii Hamiltonian
\begin{align} \label{eq:gpe}
 H_\text{GPE}[\psi] = -\frac{\hbar^2}{2m} \nabla^2 + V_0\cos^2(\kL x) + V_\text{ext}(\mathbf{r}) + \frac{4\pi\hbar^2a_s}{m} |\psi|^2,
\end{align}
where the lattice depth is typically $V_0 \approx 10\,\Er$ ($13\,$kHz), with the recoil energy defined as $\Er = (\hbar \kL)^2 / (2m)$ for a cesium atom with mass $m$. 

In addition to the lattice potential, we levitate the atoms against gravity by a magnetic field gradient~\cite{kraemer2004, dicarli2019}, and apply additional confinement using far-detuned laser beams $D_1$ and $D_2$ with global trap frequencies $\omega_{x,y,z}=2\pi\times(26, 29, 13)$\,Hz. The interaction strength is controlled by tuning the $s$-wave scattering length $a_s$ via a broad magnetic Feshbach resonance with a zero crossing at $17.1\,$G~\cite{gustavsson2008}. For detection, we use absorption images that show the atoms' momentum distribution [Fig.~\ref{Fig:Setup}(c)], after an expansion time of 72\,ms with near zero scattering length, or show their spatial distribution for a short expansion time of 2\,ms. Further details of the experimental setup are provided in~\cite[Appendix \ref{sec:Details_Setup}]{dicarli2019}.

\subsection{Periodically driven BEC \label{sec:Exp_Driving}}

We subject our system to a periodic driving force $\FD(t)$ implemented by modulating the frequency difference between the lattice beams using an acousto-optic modulator~\cite{bendahan1996a,lignier2007,sias2008,parker2013,ha2015,dicarli2023}. This modulation controls the lattice velocity $v_\text{L}(t)$ and thereby induces an inertial force in the lattice frame [Fig.~\ref{Fig:Setup}(b)]. For weak driving strengths, where interband excitations are negligible, the frequency modulation maps directly to the atomic micromotion in quasimomentum space, $\hbar k(t) = \hbar k_0 - m v_\text{L}(t)$~\cite{arimondo2012a}, with the initial momentum $\hbar k_0$ of the superfluid at the start of the drive at $t=0$.

For a periodic modulation with micromotion $k(t)=k(t+T)$ and period $T=2\pi/\omgd$, the effective single-particle dispersion $\bar{E}$ is obtained by time-averaging the band energy $E(k) = -2J \cos(k\dL)$ as~\cite{eckardt2017} 
\begin{align}\label{eq:AvEnergy_Medium}
    \bar{E}(k_0,K) = \overline{E[k(t)]} = -2 |\Jeff| \cos\!\big(k_0 \dL - \Phi\big).
\end{align}
Here, $\Jeff$ is a complex, drive-renormalized tunneling amplitude that depends on properties of the drive, such as the dimensionless driving strength $K$, and on the normal tunneling amplitude $J$ in the lattice potential. This dispersion relation can be shifted by a Peierls phase factor $\Phi = \arg(\Jeff)$ for a wave packet with momentum $\hbar k_0$ in a stroboscopic description. The realization of $\Phi$ for pulsed and multi-tone drives is discussed in Sec.\,\ref{sec:PeierlsPhase}.

We confirmed the micromotion in momentum space for both driving schemes by measuring the center-of-mass momentum. Since the momentum $\hbar k(t)$ is defined in the shaken lattice frame, whereas our absorption images are taken in the laboratory frame, an additional transformation step is required (see Appendix \ref{sec:Reference_Frames}). When measuring the real momentum distribution in the laboratory frame as in Fig.\,\ref{Fig:Setup}(c), the peak position indicates the initial momentum $\hbar k_0$, while the shift of the center-of-mass momentum provides the micromotion over one driving period. Note that additional non-inertial forces in the laboratory frame, e.g., due to gravity or magnetic field gradients, can also lead to a shift of the interference peaks.

To realize non-zero values of the initial momentum $\hbar k_0$, we applied a short acceleration pulse by reducing the levitation gradient by a few percent over a $5\,$ms interval. The pulse parameters were calibrated to produce the desired momentum distribution in a non-driven setting. The control of $k_0$ is especially important when studying parameter regimes where $\Jeff(K)$ is negative or complex, causing the system to become unstable (see Sect.~\ref{sec:Exp_PhononModes}). In this work, the system was always initialized at $t=0$ in the effective ground state of the driven system, either at $k_0 = 0$ or at $k_0 = \kL$ [Fig.~\ref{Fig:Setup}(c)], except for the measurements of the phase $\Phi$ in Sec.\,\ref{sec:PeierlsPhase}. In the figures, red lines indicate the critical driving strengths $K_\text{GS}$ at which $\Jeff$ flips sign and $k_0$ changes [e.g.~Fig.~\ref{Fig:TwoFreq_MeasureJeff}(a)].


\subsection{Phonon excitations \label{sec:Exp_PhononModes}}

In the weakly-interacting superfluid regime, excitations with momentum $q$ above the BEC mode with momentum $k$ are described in momentum space by the Bogoliubov Hamiltonian~\cite{lellouch2017}
\begin{equation}
    H_\text{BdG} =  E_0 + \sum_{q\neq k} \left[\xi(k,q) {+} g \right] a^\dagger_q a_q + g\sum_{q \neq k} (a^\dagger_q a^\dagger_{-q} {+} \mathrm{h.c.}),
    \label{eq:H_BdG}
\end{equation}
where $g=4\pi\hbar^2a_s\, \rho/m$ with $\rho$ the atom density in the condensate, and $\xi(k,q)=E(k+q)-E(k)$; 
the term $E(k)$ arises from the chemical potential and ensures the quadratic expansion around the lowest-energy BEC state with energy $E_0$. The interaction strength $g$ controls the amplitude of exciting two phonons with opposite momenta $\pm q$. Diagonalizing this quadratic model gives rise to the phonon dispersion relation $E_\text{ph}(k,q)$, see~Appendix \ref{app:theory}.

Our system exhibits several intrinsic mechanisms that give rise to phonon excitations. Even in the absence of external driving, Landau instabilities and negative-mass instabilities can occur at particular momenta of the condensate. Landau instabilities occur when $E_\text{ph}(k,q)<E_0$, i.e., the phonon energy becomes smaller than the condensate energy, indicating that creation of phonon modes allows the system to lower its energy. In contrast, negative-mass instabilities appear for $k > \kL/2$, where the effective mass is negative and phonon energies become complex. In this regime, the imaginary part of the phonon dispersion $E_\text{ph}$, determines the growth rate $\Gamma_q$ of the phonon mode, while the real part governs its phase oscillation. Notably, they conserve both the total energy and momentum of the system when phonon modes grow in pairs with opposite momenta~\cite{lellouch2017}. While Landau instabilities typically become sizable on time scales that exceed our observation window~\cite{wu2003}, negative-mass instabilities lead to significant phonon growth within observable durations~\cite{cristiani2004,fallani2004}. We adopted the term ``negative-mass instability'' in place of the common ``dynamical instability'' (exponential growth) or ``modulational instability'' (density modulation)~\cite{wu2003, fallani2004} to avoid confusion with other instabilities.

Introducing a periodic driving force is tantamount to replacing $k{\to} k(t)$ in Eq.~\eqref{eq:H_BdG}, and unlocks another class of instabilities in the system -- parametric instabilities. These arise from the time-periodic modulation of the phonon energy $E_\text{ph}[k(t),q]$, caused by the micromotion $k(t)$ of the medium. Phonon modes undergo dynamical growth when the driving frequency is within the Bogoliubov band width, $\hbar\omgd<\Delta W$, and resonantly matches the renormalized phonon energy $E^{av}_\text{ph}(k_0 ,q ; K)$ of the time-averaged Hamiltonian~(Appendix~\ref{app:theory}). They grow in pairs conserving momentum, but unlike negative-mass instabilities, their total energy is positive requiring external input from the drive. Parametric resonances in lattice systems with a single-frequency driving force are crucial for Floquet-engineered bosonic systems, and have been investigated both theoretically~\cite{creffield2009,bukov2012, lellouch2017, lellouch2018} and experimentally~\cite{boulier2019, wintersperger2020, dicarli2023, cruickshank2024}.
 
The growth rate $\Gamma_q$ of a phonon with momentum $q$ under parametric instabilities can be estimated from a mapping to an effective parametric oscillator (see~Appendix~\ref{app:theory}). Importantly, both the magnitude of the instability and the width of the parametric resonance scale proportionally with the absolute value of the resonant Fourier coefficient of the time-periodic function~\cite{lellouch2017}
\begin{equation}
    h_{q}(t) = \frac{1}{2}\left(\xi[k(t),q] + \xi[k(t),-q]\right) = \sum_{n} h_{q,n}\; e^{-i n\omgd t},
\label{eq:hq(t)}
\end{equation}
i.e. $\Gamma_q{\propto}|h_{q,n}|$ where $n\omgd\approx 2 E^{av}_\text{ph}(k_0,q;K)$ and $n{\neq}0$. The experimentally measurable total growth rate $\Gamma$ is determined by averaging over all momentum modes, weighted by their initial population; at long times, it is dominated by $\Gamma_\infty = \max_q \Gamma_q$, corresponding to the maximally unstable mode $q_\text{mum}$~\cite{lellouch2017}.  

Parametric and negative-mass instabilities dominate the system in different regimes of the driving frequency $\omgd$ and driving amplitude $K$. In the limits of very slow driving ($\hbar \omgd \ll J$) and fast driving ($\hbar \omgd > \Delta W$), negative-mass instabilities prevail. In these regimes, the system becomes unstable whenever either the bare dispersion $E_\text{ph}(k, q)$ or the renormalized dispersion $E^{av}_\text{ph}(k_0, q; K)$ develops an imaginary component. Parametric resonances, on the other hand, occur in the intermediate regime $\hbar \omgd \sim \Delta W$. To isolate the effects of parametric from negative-mass instabilities in this regime, we restricted the driving strength to values below a critical threshold $K_c$, such that $k(t) < \kL/2$ at all times and the system never enters the negative-effective-mass region. In the figures, this critical value is indicated by white lines labeled by $K_c$ [e.g., see Fig.\,\ref{Fig:TwoFreq_MeasureJeff}(a)].

To measure the total phonon growth rate $\Gamma$, we prepared the system with an initial momentum $k_0$ corresponding to the ground state of the time-averaged energy, $\bar{E}(k_0,K)$. The system is then driven for a duration $t$, chosen as an integer multiple of the driving period $\TD$, after which the real momentum distribution of the atoms is measured in the laboratory frame using absorption imaging. To avoid complications in interpreting the final momentum distribution, we simultaneously switch off both the drive and the lattice beams, deliberately avoiding band-mapping techniques that rely on ramping down the lattice to reveal the quasimomentum distribution. The relative number of excited atoms, $\Nex/\Ntot$, was obtained by counting atoms within the regions marked by red patches in Fig.\,\ref{Fig:Setup}(c). These regions exclude momenta near $0,\pm2\hbar\kL$ for $k_0=0$, and near $\pm\hbar \kL,\pm3 \hbar \kL$ for $k_0=\kL$. This approach does not account for phonon modes with small values of $q$, which are, however, expected to contribute negligibly due to their low growth rates.

\section{Pulsed drive \label{sec:Pulsed_Drive}}

For the pulsed driving scheme, we implement a micromotion with momentum
\begin{align}\label{eq:Pulsed_Micromotion}
    \hbar k_\text{pul}(t) =\hbar  k_0 + K \frac{4\hbar }{\pi\dL} \sum_{n=1,3,...}^{M}\frac{1}{n}\sin\left(n\omgd t\right),
\end{align}
where $M$ indicates the number of odd harmonics of the wave form. For increasing $M$, the micromotion approaches a square wave, while the corresponding force shows sharp, alternating pulses [Fig.~\ref{Fig:Setup}(b)]. The prefactor $4/(\pi\dL)$ in Eq.\,(\ref{eq:Pulsed_Micromotion}) is chosen such that a dimensionless driving strength of $K=\pi$ creates a micromotion that reaches the edge of the Brillouin zone in the limit of large $M$~\footnote{Note that this definition of $K$ varies by factor $4/\pi$ with respect to Refs.~\cite{dicarli2023, cruickshank2024}.}. We experimentally verified this implementation of the micromotion in momentum space by measuring the center-of-mass momentum over a full driving period (see Appendix~\ref{sec:Reference_Frames}).

\subsection{\label{subsec:pulsed-BEC}BEC mode subject to pulsed driving}

Time-averaging the energy of the BEC mode for this micromotion provides an effective dispersion relation with a renormalized tunneling $\Jeff(K,M)$ that depends on $K$ and $M$. Simple analytical expressions for $\Jeff$ are available in the limits $M=1$ and $\infty$. For a single harmonic, $\Jeff$ is given by the zeroth-order Bessel function $\Jeff = J\mathcal{J}_0\left(4K/\pi\right)$. In the limit of many harmonics with a micromotion of the form $k_\text{PD}(t) = k_0 + \left(K/\dL\right)\text{sign}[\sin(\omgd t)]$, the effective tunneling simplifies to $\Jeff = J\cos(K)$. The same approach can be applied to recover the renormalized phonon energy, $E^{av}_\text{ph}$ (see Appendix\,\ref{app:theory}).

\begin{figure} 
\includegraphics[width=\columnwidth]{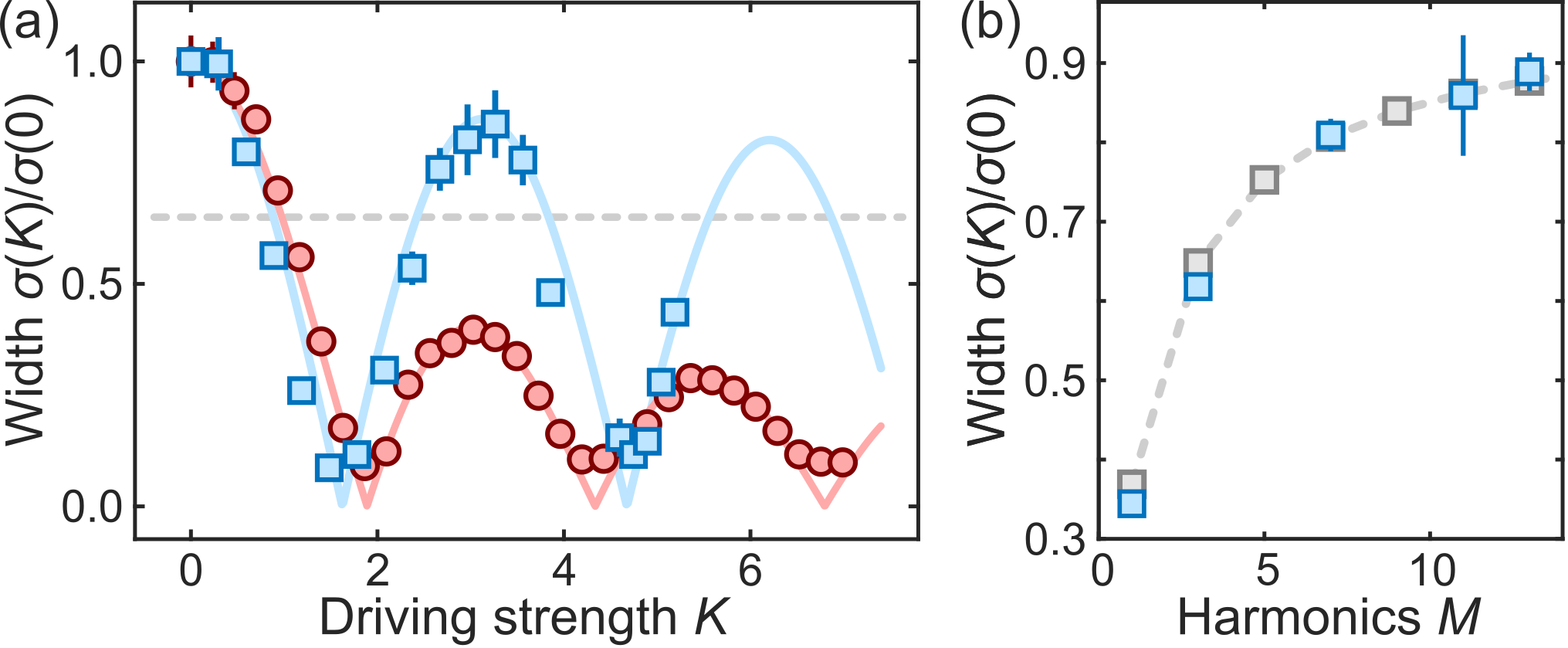}
\caption{Pulsed driving. (a) Measurement of the normalized cloud width to determine $|\Jeff/J|$ for sinusoidal drive (red circles, $t=600\,$ms, $V_0=11\,\Er$) and for pulsed drive (blue squares, $t=800\,$ms, $V_0=8.3\,\Er$, $M=11$, $\TD=6\,$ms, $a_s=2\,a_0$). Solid lines denote the calculated tunneling energy $\Jeff/J$. The horizontal gray dashed line indicates $|\Jeff/J|=0.65$. (b) Measured cloud width for drive of varying harmonics $M$ at fixed $K=3.26$ (blue squares) compared to numerically calculated $\Jeff/J$ (gray squares). The dashed gray line is a guide to the eye. The strong agreement demonstrates that pulsed driving enables a precise control of $|\Jeff/J|$ using both $K$ and $M$.}\label{Fig:SquareDrive_Jeff}
\end{figure}

In a first measurement, we experimentally validated the predicted tunneling amplitude $\Jeff$ for pulsed driving with weak interaction by measuring the spatial spreading of the atomic cloud after hold times of hundreds of milliseconds~\cite{lignier2007}. The resulting absorption images show a strong variation of the spreading in position space as a function of $K$ and $M$ [Fig.\,\ref{Fig:Pulsed_MeasureJeff_A}(b)]. To quantify this spreading, we measured the cloud's width after time $t$ and normalized to the value without driving, $\sigma(K)/\sigma(0)$. This normalized width is expected to reflect the magnitude of $|\Jeff/J|$ (see Appendix\,\ref{sec:Pulsed_MeasureJeff_A}). 

Our measurements show excellent agreement with numerical calculations, without the use of any free parameters. This includes the recovery of the expected Bessel-function dependence for sinusoidal driving, $M=1$ [red circles in Fig.\,\ref{Fig:SquareDrive_Jeff}(a)]~\cite{lignier2007}. For pulsed driving with a finite number of harmonics, $M=11$ [blue squares in Fig.\,\ref{Fig:SquareDrive_Jeff}(a)], the measured dependence already approaches the cosine behavior expected for large values of $M$. Solid lines in Fig.\,\ref{Fig:SquareDrive_Jeff}(a) show the corresponding calculated values of $\Jeff/J$. Varying the number of harmonics $M$ allows for further tuning of $\Jeff$, and our measurement data [blue squares in Fig.\,\ref{Fig:SquareDrive_Jeff}(b)] show again good agreement with the numerically calculated values of $\Jeff/J$ (gray squares).

\subsection{Instabilities for pulsed driving \label{sec:Pulsed_Phonons}}

A key finding of our work is that pulsed driving schemes offer a versatile new approach to avoid phonon excitations resulting from both effective-mass instabilities and parametric instabilities. To demonstrate this reduction of the phonon growth rate $\Gamma$ for pulsed driving compared to sinusoidal driving, we chose driving strengths for both cases with identical effective engineered tunneling $|\Jeff/J|=0.65$ [dashed line in Fig.\,\ref{Fig:SquareDrive_Jeff}(a)]. This corresponds to a driving strength of $K=0.9$ for pulsed and $K=1.0$ for sinusoidal driving. Both driving strengths are sufficiently small to avoid negative-mass instabilities, which occur at $\Kc=\pi/2$ and $\pi^2/8$ for pulsed and sinusoidal drives, respectively. After a variable driving duration $t$, we measured the fraction of excited atoms in momentum space, $\Nex/\Ntot$.

\begin{figure} 
\includegraphics[width=\columnwidth]{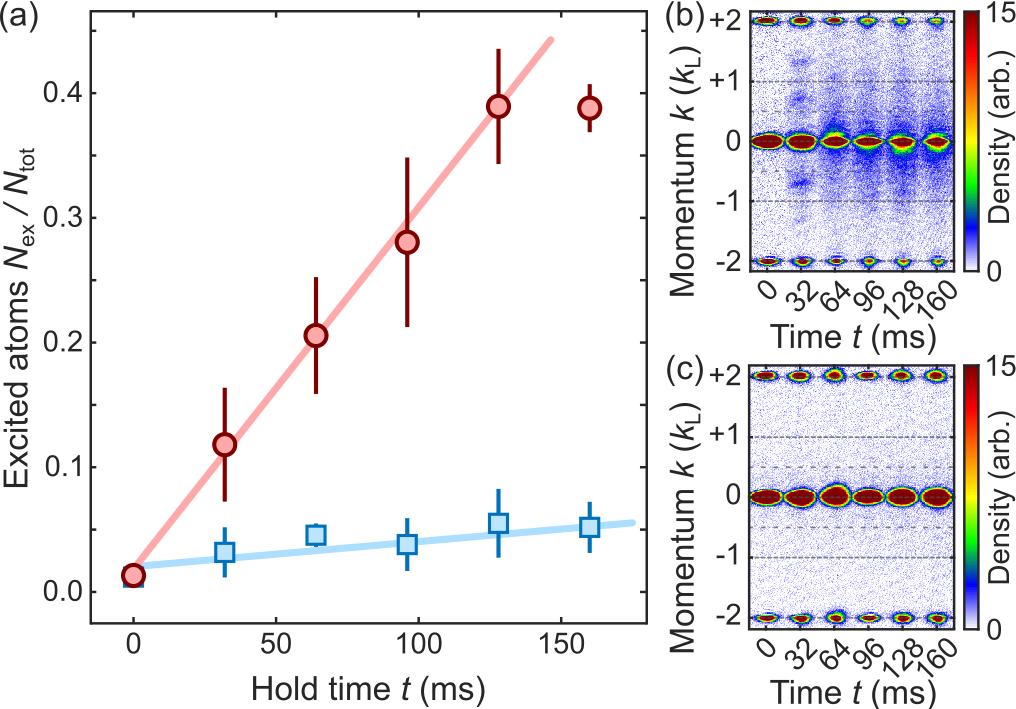}
\caption{Suppression of phonon growth for pulsed driving. (a) Fraction of atoms in phonon modes for sinusoidal driving (red circles, $K=1.0$) and pulsed driving (blue squares, $K=0.9$, $M=11$), both with parameters $|\Jeff/J|=0.65$, $\TD=4\,$ms, $V_0=10.3\,\Er$, $a_\text{s}=88\,a_0$. Solid lines are linear fits to the data. (b) Absorption images for sinusoidal driving $\left(M=1\right)$ and (c) pulsed driving with $M=11$, parameters as in (a). For pulsed driving, phonon modes are no longer visible, indicating a suppression of $\Gamma$ by at least a factor of eight compared to the sinusoidal drive.}\label{Fig:Pulsed_MeasurePhonons}
\end{figure}

For sinusoidal driving [red circles in Fig.\,\ref{Fig:Pulsed_MeasurePhonons}(a)], we observed pronounced phonon growth, as discussed in Refs.\,\cite{boulier2019, wintersperger2020, dicarli2023, cruickshank2024}. Initially, phonon excitations appear at well-defined quasimomenta [$t=32\,$ms in Fig.\,\ref{Fig:Pulsed_MeasurePhonons}(b)], but over time they couple to additional modes and spread across the Brillouin zone. The number of atoms in unstable modes grows approximately linearly until reaching 40\% of the total atom number [red circles in Fig.\,\ref{Fig:Pulsed_MeasurePhonons}(a)]. Remarkably, pulsed driving almost completely suppresses the growth of phonon modes [blue squares in Fig.\,\ref{Fig:Pulsed_MeasurePhonons}(a)]. No clear indication of unstable modes was observed within 160\,ms (forty driving cycles), and $\Nex/\Ntot$ remained at the noise floor of our imaging system. The effect is striking when comparing the absorption images of the two driving schemes [Fig.\,\ref{Fig:Pulsed_MeasurePhonons}(b) and (c)].

An intuitive explanation for the suppression of parametric instabilities is given by the square-like shape of the micromotion. Parametric instabilities arise from the periodic modulation of system parameters, induced by the micromotion. In the case of a square micromotion, however, $k(t)$  switches abruptly between two fixed values, spending little time in the intermediate regime. This reduces modulations and the resulting growth of phonon modes. Formally, one can understand the mechanism behind this suppression by recalling the parametric instability rate $\Gamma\propto |h_{q,n}|$, see Eq.~\eqref{eq:hq(t)}. A straightforward calculation using the pulsed-driven free dispersion $E[k(t),q]$ shows that $h_{q,n}{\equiv}0$ for $n{\neq}0$ (see~Appendix~\ref{sec:A_Pulsed_Theory}). As a result, we find that the rate $\Gamma{\equiv}0$ vanishes identically for a perfect pulsed drive. In the experiment, we consider an approximation consisting of $M$ harmonics; therefore, the instability rate is not perfectly inhibited, but still substantially reduced. 

In addition, pulsed driving automatically suppresses negative-mass instabilities. For sinusoidal driving with $k_0=0$, they arise once the driving strength exceeds a critical threshold $K>K_c$, causing the micromotion to enter regions of the Brillouin zone with negative effective mass. This occurs already for values of $K$ below $K_\text{GS}$, the threshold at which the initial momentum switches for the ground state from $\hbar k_0=0$ to $\hbar k_0=\hbar \kL$. For pulsed driving, however, $K_\text{GS}$ approaches $K_c$ as the number of harmonics $M$ increases, reaching the limit $\Kc=K_\text{GS}=\pi/2$. As a result, $k_0$ changes precisely when the micromotion approaches the negative-effective-mass regions, ensuring that $k(t)<0.5\kL$ at all times. Pulsed driving therefore simply removes the entire interval of $K$-values for which negative-mass instabilities would otherwise occur (see Appendix~\ref{sec:Pulsed_Motivation}).

While the pulsed driving scheme effectively suppresses phonon excitations arising from parametric and negative-mass instabilities, it introduces a technical limitation for large values of~$M$. The short pulses must be sufficiently weak to prevent Landau-Zener transitions to higher bands~\cite{weinberg2015multiphoton}. The peak acceleration in our driving scheme increases with $K$, $\omgd$, and $M$, while the critical acceleration for Landau--Zener tunneling depends on the lattice depth~$V_0$~\cite{jonalasinio2003}. We find that band excitations occur in our measurements for specific combinations of parameters, typically showing as a loss of atoms in absorption images. For instance, this behavior is visible for driving strengths near $K=4$ in Fig.\,\ref{Fig:Pulsed_MeasureJeff_A}(b). We excluded those data sets from the measurement of $\Jeff$.

\section{Two-tone drive \label{sec:TwoFreq_Drive}}

\begin{figure} 
\includegraphics[width=\columnwidth]{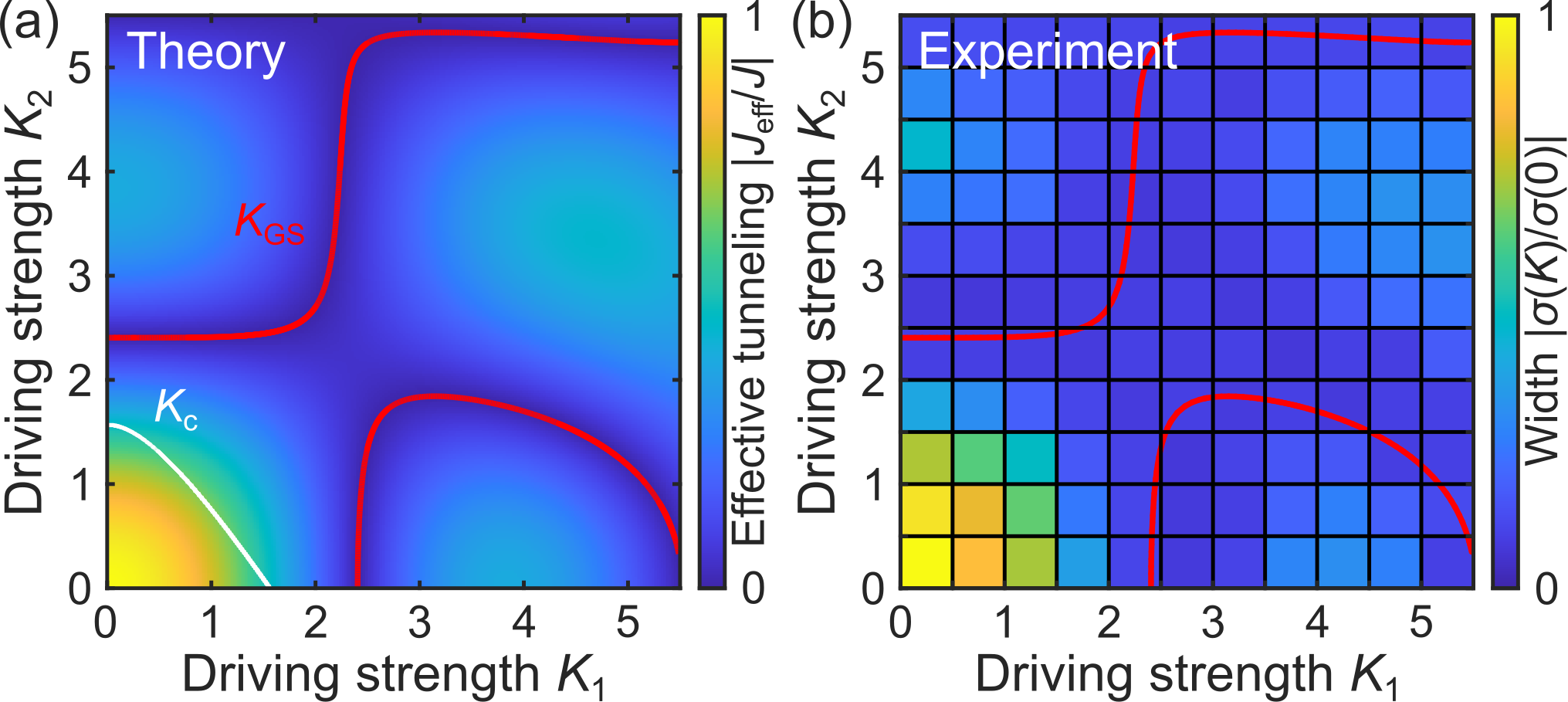}
\caption{Tunneling for two-tone driving. (a) Calculated magnitude $|\Jeff(K_1,K_2,0)/J|$ from Eq.\,(\ref{eq:TwoFreq_Dispersion}). Red contours indicate $\Jeff=0$. The white curve marks the minimum driving strengths required for negative-mass instabilities. (b) Measured BEC cloud width $\sigma(K)/\sigma(0)$ after $t=600\,$ms for two-tone modulation with $\TD=4\,$ms, $V_0=10\,\Er$, $\varphi=0$, and $a_\mathrm{s}=2\,a_0$. The measured width is proportional to $\Jeff/J$ and is in good agreement with the theoretical prediction.} \label{Fig:TwoFreq_MeasureJeff}
\end{figure}

The ability of the pulsed drive to suppress (and even fully inhibit, for $M\to\infty$) parametric instabilities relies on the particular values for its Fourier coefficients. Motivated by the underlying mechanism, we now show that it is possible to design periodic drives that suppress parametric instabilities by carefully selecting their Fourier content. 

In particular, we now demonstrate that a significant reduction of $\Gamma$ can already be achieved using just two driving frequencies with the ratio $\omega_2 = 2\omega_1$ and driving strengths $K_1$ and $K_2$. We define these strengths such that the micromotion for two-tone drive (tt) is given by
\begin{align}\label{eq:micromotion2}
\hbar k_\text{tt}(t) = \hbar k_0  + \frac{\hbar\kL}{\pi} \left[ K_1 \sin(\omega_1 t) + K_2 \sin(\omega_2 t + \varphi) \right],
\end{align}
with the micromotion reaching the edge of the Brillouin zone, $k_\text{tt} = \kL$, for the cases $(K_1, K_2) = (\pi, 0)$ and $(0, \pi)$. This definition closely matches our experimental implementation, which directly controls the micromotion rather than the driving force. As a result, equal driving strengths, $K_1 = K_2$, correspond to forces with amplitudes differing by a factor two. The parameter $\varphi$ describes the relative phase between the two driving frequencies and is directly observable in the experiment through the micromotion in momentum space.

Generalizing the results from Sec.~\ref{sec:Exp_Driving}, we integrated the micromotion over one driving period to obtain the effective dispersion at lowest order, $\bar{E}(k_0,K_1,K_2,\varphi)$, which amounts to a renormalization of the tunneling amplitude into the effective tunneling amplitude
\begin{align}\label{eq:TwoFreq_Dispersion}
    \Jeff(K_1,K_2,\varphi) = J \sum_{n=-\infty}^\infty \mathcal{J}_{-2n}(K_1)\mathcal{J}_{n}(K_2) e^{in\varphi},
\end{align}
where the Peierls phase factor $\Phi=\arg[\Jeff]$ depends also on the relative phase $\varphi$. In general, computing the full effective Hamiltonian would include higher-order terms featuring longer-range hopping. Figure \ref{Fig:TwoFreq_MeasureJeff}(a) shows the calculated values of $\Jeff/J$ for $\varphi=0$. Note that the critical value $K_\text{GS}$, where the ground state changes from $k_0=0$ to $\kL$ at the zero crossing of $\Jeff$, is no longer a single value but a line in the ($K_1,K_2$)-plane [red line in Fig.\,\ref{Fig:TwoFreq_MeasureJeff}(a)]. The same is true for the threshold value $K_c$, where the micromotion reaches $k_\text{tt}(t)=0.5\kL$ and negative-mass instabilities occur [white line in Fig.\,\ref{Fig:TwoFreq_MeasureJeff}(a)].

Similar to the case of pulsed driving (Sec.\,\ref{sec:Pulsed_Drive}), we validated Eq.\,(\ref{eq:TwoFreq_Dispersion}) by studying the spreading of a wave packet under a two-tone drive. For $\varphi=0$, the relative width $\sigma(K)/\sigma(0)$ of the expanding wave packet provides a measure of $|\Jeff/J|$. Using the procedure described in Appendix\,\ref{sec:Pulsed_MeasureJeff_A}, we measured the relative width after $t = 600$\,ms [Fig.\,\ref{Fig:TwoFreq_MeasureJeff}(b)] and compared it to the  predicted renormalized tunneling $|\Jeff(K_1, K_2, \varphi=0)/J|$ [Fig.\,\ref{Fig:TwoFreq_MeasureJeff}(a)]. Again, red lines indicate a changing ground state from $k_0=0$ to $\kL$. We find good agreement between the measured values of $\Jeff$ and the theoretical prediction.

\subsection{Instabilities for two-tone drive}
\begin{figure}
\includegraphics[width=\columnwidth]{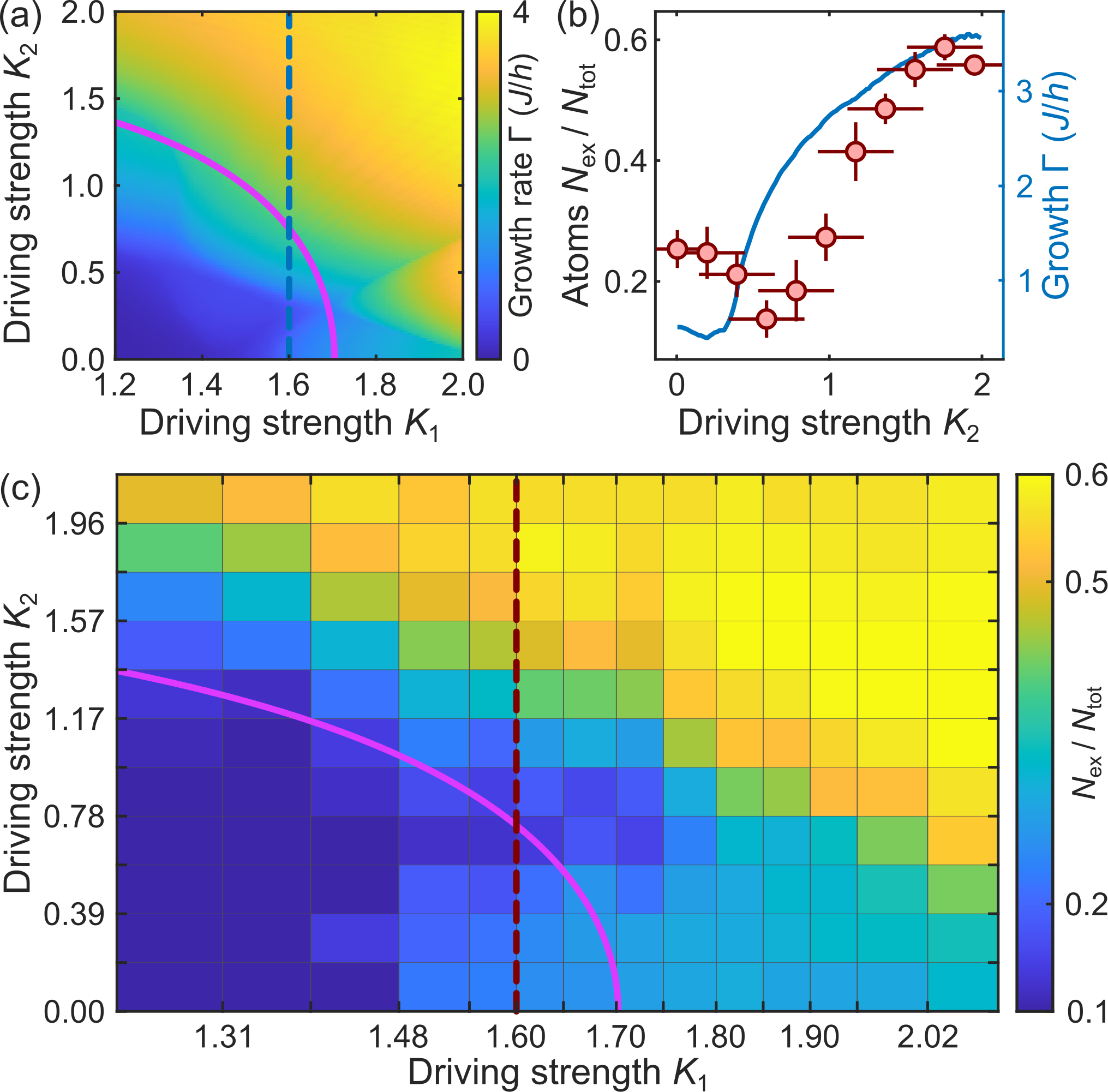}
\caption{Phonon growth for two-tone drive. (a) Numerically calculated growth rate $\Gamma$ for $g/J=50$, $T_1=2\,T_2=6\,$ms, $V_0=10.3\,\Er$. The solid purple line marks driving parameters with $|\Jeff/J|=0.4$, while the dashed blue line indicates the driving strengths used in (b). (b) Measured fraction of atoms in unstable modes after $18\,$ms for $K_1=1.6$ (red circles), compared to the simulated $\Gamma$ from (a) (blue line). (c) Measured growth rate for two-tone driving after after $18\,$ms with $V_0=10.3\,\Er$, $T_1=2\,T_2=6\,$ms, $a_\text{s}=+88\,a_0$. The red dashed line denotes the data points plotted in (b). The figure demonstrates the reduction of instabilities by tuning the strength of the second driving component $K_2$. \label{Fig:TwoFreq_PhononGrowth} }
\end{figure}

The addition of a second driving frequency increases the parameter space of the system and allows for greater and more universal control of both $|\Jeff/J|$ and the instability rate $\Gamma$. To simulate the phonon growth for this driving scheme in $(K_1,K_2)$-space, we calculated $\Gamma$ using the BdG equations associated with Eq.~\eqref{eq:H_BdG}, and averaged over fifty-one evenly spaced phonon modes with $q$ between $0$ and $\pi$ [Fig.\,\ref{Fig:TwoFreq_PhononGrowth}(a)]. For reference, we provide a line with constant $|\Jeff/J|$ that forms a quarter-circle for $K_1,K_2<2.4$ [purple line in Fig.\,\ref{Fig:TwoFreq_PhononGrowth}(a)]. The growth rate strongly varies along this contour, allowing us to increase the stability of an initially unstable system with $(K_1 \ne 0,K_2 = 0)$ by introducing a nonzero value of $K_2$, while keeping $|\Jeff/J|$ constant.

To experimentally demonstrate the variation of $\Gamma$, we measured the stability of a system with varying strengths $K_1$ and $K_2$, where $\omega_2 = 2\omega_1$. The observed fraction of excited atoms $N_\text{ex}/N_\text{tot}$ after $18\,$ms was again used as an indicator of the phonon growth rate, showing stable and unstable regions in the resulting stability diagram with blue and yellow colors, respectively [Fig.\,\ref{Fig:TwoFreq_PhononGrowth}(c)]. For strong driving ($K_1 > 1.8$), introducing the second drive with amplitude $K_2$ generally reduced the stability [yellow region in Fig.\,\ref{Fig:TwoFreq_PhononGrowth}(c)]. In contrast, for intermediate driving strengths ($K_1 \approx 1.6$), the addition of the second frequency can increase the stability. For example, in the case where $|\Jeff/J| = 0.4$ [purple line in Fig.\,\ref{Fig:TwoFreq_PhononGrowth}(c)], our measurement showed an excited fraction of 30$\%$ after three driving cycles for $K_1=1.7$. With optimized values of $K_1\approx1.3$ and nonzero $K_2\approx1.2$, this was reduced to 12$\%$.  
  
Alternatively, increasing $K_2$ while keeping $K_1$ fixed can also reduce the phonon growth rate, as indicated by the dashed red line in Fig.\,\ref{Fig:TwoFreq_PhononGrowth}(c). Measuring phonon growth along this trajectory with $K_1 = 1.6$ [Fig.\,\ref{Fig:TwoFreq_PhononGrowth}(b)] reveals a clear suppression of instability around $K_2 \approx 0.8$, followed by a steady increase in phonon growth at larger values of $K_2$. Here, the error bars denote the width of the superfluid in momentum space, which broadens the amplitude of the micromotion. This approach of instability reduction is particularly useful in scenarios where increasing the driving frequency to reach the fast driving regime is impractical, e.g., such as in systems with small band-gaps or with tilted potentials~\cite{hui2010, lellouch2018,cruickshank2024}. Additional discussion of the pattern in Fig.\,\ref{Fig:TwoFreq_PhononGrowth}(a) that enables the reduction of phonon growth with two-tone drive is provided in Appendix \ref{sec:TwoFreq_Motivation}. We note a shift in the observed structure along the $K_2$-axis between experimental measurements and numerical simulations [see Fig.\,\ref{Fig:TwoFreq_PhononGrowth}(b)]. We expect this to be caused by a change in the atomic density spatially along the lattice direction due to the harmonic confinement, which is unaccounted for in the simulations. In general, however, we find good qualitative agreement between our measurement results in Fig.\,\ref{Fig:TwoFreq_PhononGrowth} and the simulation in Fig.\,\ref{Fig:TwoFreq_PhononGrowth}(a).

\section{Peierls phase} \label{sec:PeierlsPhase}

\begin{figure} [t]
\includegraphics[width=\columnwidth]{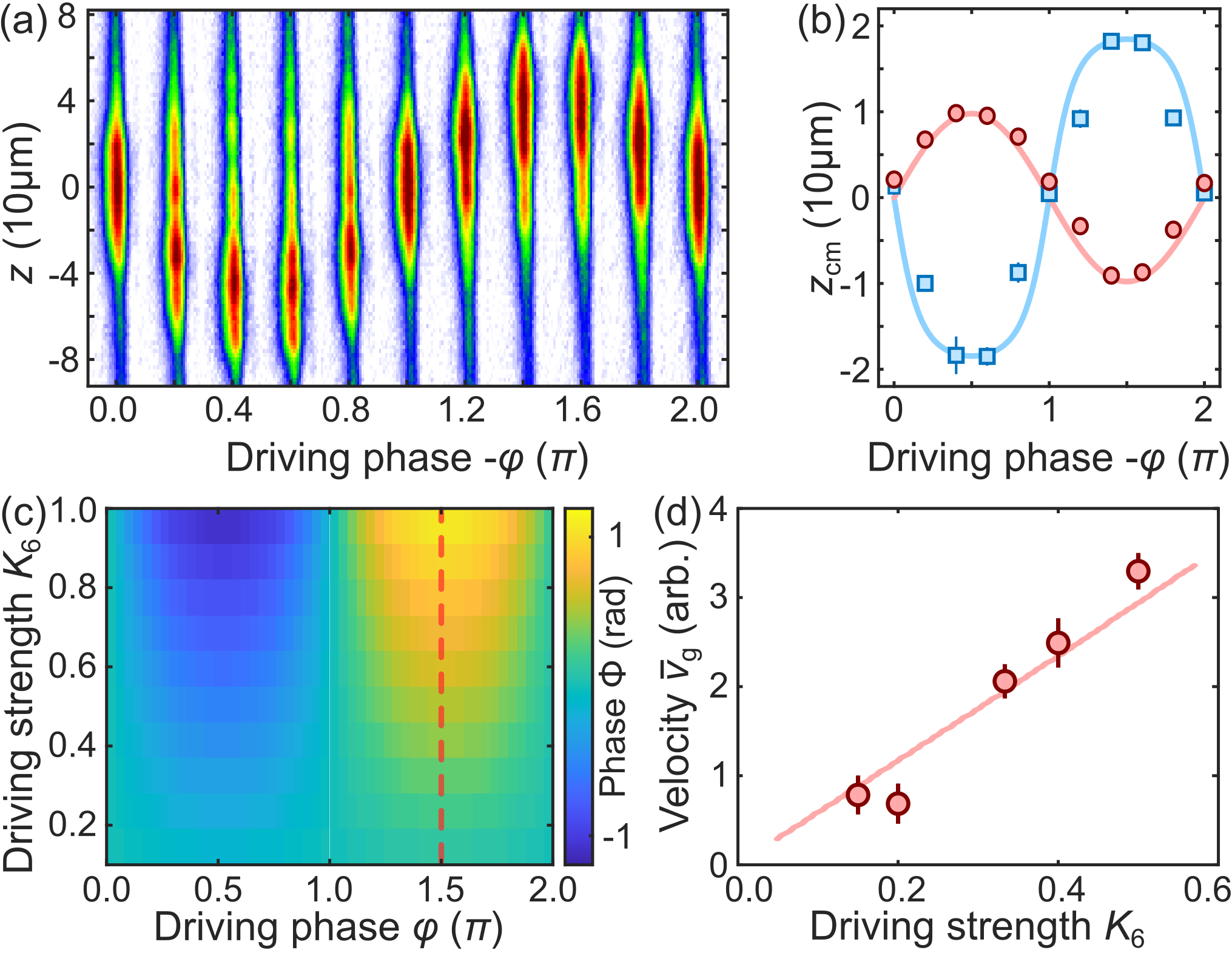}
\caption{Peierls phase factor and resulting motion for our driving schemes [Eq.\,\eqref{eq:micromotion2} and Eq.\,\eqref{eq:MicromotionPulsedBroken}]. (a) Absorption images in position space after $t=150$\,ms for varying phase $\varphi$ with $K_1=1.1$, $K_2=3.6$, $\TD=3\,$ms, $V_0=4\,\Er$, $a_s=2\,a_0$, $k_0=\kL$. (b) Measured center-of-mass position $z_{cm}$ for data in (a) (blue squares) and for $K_2=2.2, k_0=0$ (red circles). Lines are calculated using the group velocity in Eq.\,\eqref{eq:qroupvelocity} and a free fitting parameter for the amplitude. (c) Calculated Peierls phase for pulsed driving with an additional driving term with strength $K_6$ and phase $\varphi$. Dashed red line denotes scan parameters in (d). (d) Measured center-of-mass shift of the cloud over 150\,ms for $\varphi=3/2\pi$ and varying strength $K_6$, parameters $T_\text{D}=6\,$ms, $V_0=4\,\Er$, $M=11$. The red line was calculated using Eq.\,\eqref{eq:qroupvelocity} with a free fitting parameter for the amplitude. Both driving schemes can be utilized to introduce a controllable Peierls phase to the system.}\label{Fig:Motion}
\end{figure}

Complex tunneling of the form $\Jeff = |\Jeff| e^{i\Phi}$ with a Peierls phase $\Phi$ has been realized in numerous experiments using periodic driving~\cite{alberti2009, haller2010a, aidelsburger2011, aidelsburger2013, aidelsburger2015, struck2012a, miyake2013, grossert2016, tai2017}. In our 1D system, such complex phases are associated with an overall momentum kick, and hence, they do not affect the ground-state properties; in particular, this effect can be eliminated by transforming to an appropriate moving frame. More generally, in the Floquet framework, such gauge transformations can be implemented through a unitary, or ``kick'', operator~\cite{goldman2014}. However, Peierls phase factors can lead to non-trivial physical effects when they correspond to a non-zero flux. In 1D, this can occur for a closed chain~\cite{amico2022}, while in 2D, Peierls phase factors can generate a non-zero flux through the lattice, giving rise to Landau-level-type physics~\cite{goldman2014,cooper2019topological}, quantum Hall effects~\cite{leonard2023realization}, and quantized vortices~\cite{lin2009}.

Here, we demonstrate that $\Phi$ can be easily incorporated into the renormalized dispersion relation Eq.\,\eqref{eq:AvEnergy_Medium} for our two driving schemes. A directed current emerges whenever the driving protocol breaks reflection and shifts symmetries simultaneously~\cite{Flach2000,heimsoth2010,creffield2011,struck2012a}. In our system, this results in a spatially asymmetric micromotion, such that forward and backward trajectories no longer cancel in position space over one driving cycle. 

This mechanism is particularly intuitive in systems with two driving terms: the first induces a reflection-symmetric spatial motion, while the second one breaks this symmetry. The concept has been previously demonstrated in various systems, e.g., in the context of Bloch oscillations, where a constant force leads to symmetric oscillations, and the addition of an off-resonant driving force breaks this symmetry, resulting in super-Bloch oscillations with large amplitude~\cite{alberti2009,haller2010a}. The mechanism has also been realized using a modulation of the tunneling amplitude $J(t)$~\cite{cao2020}, of the interaction strength~\cite{diaz2013}, and the confining potential~\cite{ali2024}. 

For the two-tone driving scheme discussed in Sec.~\ref{sec:TwoFreq_Drive}, an asymmetric micromotion arises due to the relative phase $\varphi$ in Eq.~\eqref{eq:micromotion2} for $\varphi \neq 0,\pi$~\cite{denisov2007}. In contrast to frequency detuning, which induces large-amplitude oscillatory dynamics, $\varphi$ causes a constant net displacement per driving period, thereby generating a uniform directed motion. This displacement can be calculated either by integrating the instantaneous group velocity over time or by using the effective group velocity associated with the renormalized dispersion relation,
\begin{align} \label{eq:qroupvelocity}
\bar{v}_g(k_0) \sim \partial_{k_0} \bar{E} = 2 |\Jeff| \dL \sin(k_0 \dL - \Phi).
\end{align}
Measuring $\bar{v}_g$ provides a direct probe for the Peierls phase $\Phi = \arg(\Jeff)$ in Eq.~\eqref{eq:TwoFreq_Dispersion}. In particular, for $k_0 = 0$, the group velocity depends directly on $\sin(\Phi)$.

For the pulsed driving scheme in Sec.~\ref{sec:Pulsed_Drive}, we introduced an additional term that breaks the symmetry of the micromotion, leading to a total momentum modulation 
\begin{align}\label{eq:MicromotionPulsedBroken}
k(t) = k_{\mathrm{pul}}(t) + K_6 \sin(6\omgd t + \varphi).
\end{align}
Here, $k_{\mathrm{pul}}(t)$ is the micromotion for pulsed driving defined in Eq.~(\ref{eq:Pulsed_Micromotion}), while the second term represents an even harmonic of 6th order with amplitude $K_6$ and phase offset $\varphi$. To extract the Peierls phase, we numerically calculated the cycle-averaged energy $\overline{E[k(t)]}$ and compared the resulting dispersion with Eq.~(\ref{eq:AvEnergy_Medium}). The Peierls phase causes a shift of the dispersion curve, with the minimum position indicating $\Phi$. 

To measure $\Phi$ for both driving schemes, we removed the longitudinal trapping potential by switching off laser beam $D_1$ [Fig.\,\ref{Fig:Setup}(a)] and monitored the variation of the cloud position after $150\,$ms of driving. Absorption images of the atomic density distribution show a net displacement of the cloud accompanied by a slow expansion [Fig.~\ref{Fig:Motion}(a)]. The displacement was quantified by the center-of-mass position, $\zcm=\int n(z) z dz/N_\text{tot}$, which depends on the phase offset $\varphi$ for the two-tone driving scheme and on both $\varphi$ and the harmonic amplitude $K_6$ for the case of pulsed driving.

For the two-tone driving scheme, the measured phase dependence of the cloud displacement agrees well with the theoretically predicted group velocity $\bar{v}_g$, calculated using Eq.~\eqref{eq:qroupvelocity} [Fig.~\ref{Fig:Motion}(b)]. The observed displacement arises from the relation $\bar{v}_g \sim \sin(\Phi)$ together with the dependence of the Peierls phase $\Phi$ on $\varphi$ in Eq.~\eqref{eq:TwoFreq_Dispersion}. A single scaling factor was fitted to relate $\bar{v}_g$ to the measured center-of-mass shift. As expected, the direction of the observed motion reverses with variations in the modulation strength $K_2$ for the two data sets. 

For the pulsed driving scheme with the additional $K_6$ term, we calculated $\Jeff$ numerically. The resulting Peierls phase is a function of $\varphi$ and $K_6$ with a sign depending on $\varphi$ [Fig.~\ref{Fig:Motion}(c)]. We again observe good agreement between the calculated value of $\bar{v}_g$ and the measured displacement [Fig.~\ref{Fig:Motion}(d)]. For simplicity, we fixed the phase $\varphi=3/2\pi$ and measured $\zcm$ along the red dashed in Fig.~\ref{Fig:Motion}(c). We again used a scaling factor to fit $\bar{v}_g$ to the measured displacement. In this regime, the resulting Peierls phase is small, and the sinusoidal dependence of $\bar{v}_g$ on $\Phi$ in Eq.\,\eqref{eq:qroupvelocity} is effectively linear [red line in Fig.~\ref{Fig:Motion}(d)].

\section{Conclusion}\label{sec:Conclusion}
In this work, we have demonstrated that tailored multifrequency driving schemes can effectively suppress parametric and negative-mass instabilities in periodically driven systems. Through both pulsed and two-tone drive protocols, we achieved stabilizing control of the effective tunneling amplitude and the Peierls phase, while drastically reducing phonon growth rates compared to conventional sinusoidal driving. Our experimental observations, supported by numerical simulations and perturbative analytical predictions, demonstrate that parametric instabilities can be greatly suppressed or even eliminated without sacrificing the tunability of the Floquet-engineered band structures. In particular, we found that high-harmonic pulsed drives naturally suppress excitations due to their square-wave-like micromotion, and that two-tone drives offer enhanced flexibility for optimizing stability through the control of amplitude and phase relations. 
Theoretically, our work demonstrates that the vast parameter space of drive frequencies, amplitudes, and relative phases offers substantial opportunities to design Floquet-engineering drives with additional properties, such as improved stability of excitation dynamics. 
These findings provide a pathway for extending the applicability of Floquet engineering to interacting quantum systems, enabling long-lived and stable operation in regimes that were previously inaccessible due to heating and decoherence.

Our findings are directly applicable to ultracold bosonic gases with long-range interactions~\cite{su2023dipolar}, and two-dimensional systems where Floquet drives can engineer artificial magnetic fields~\cite{goldman2014b,goldman2016topological,aidelsburger2018artificial,cooper2019topological}. It is therefore important to explore the stability enhancement offered by multi-tone driving in two- and three-dimensional bosonic lattices. 
Looking ahead, we expect our proposed suppression mechanism for parametric instabilities to directly facilitate the simulation of strongly interacting topological bosonic systems beyond the Gross-Pitaevskii equation -- including bosonic Pfaffian states requiring weak (two-body) interactions for preparation~\cite{palm2025absence} and generic periodically driven models with bosonic quasiparticle excitations.

\section*{Acknowledgments}
We acknowledge support by the EPSRC through a New Investigator Grant (EP/T027789/1), the Programme Grant QQQS (EP/Y01510X/1) and the Standard Grant (UKRI2897). SL acknowledges support by the EPSRC through the Programme Grant ``Quantum Technology Hub in Sensing, Imaging and Timing'' (Grant No. EP/Z533166/1).
MB was funded by the European Union (ERC, QuSimCtrl, 101113633). 
ED acknowledges support from the SNSF project 200021\textunderscore212899, funding by the Swiss State Secretariat for Education, Research and Innovation (Contract No. UeM019-1), 
NCCR SPIN, a National Centre of Competence in Research (SNSF Grant No. 225153).
Work in Brussels was financially supported by the ERC Grants TopoCold and LATIS, the EOS project CHEQS and the Fondation ULB.
Views and opinions expressed are, however, those of the authors only and do not necessarily reflect those of the European Union or the European Research Council Executive Agency. Neither the European Union nor the granting authority can be held responsible for them. 

\section*{Author contributions}
The original concept was developed by NG, with inputs from MB and SL. RC acquired the experimental data, and RC together with EH performed the data analysis. Analytical and numerical calculations were carried out by MB and SL, who contributed equally, with inputs from NG and ED. All authors contributed to the interpretation of the results and to the writing of the manuscript.


\renewcommand{\thefigure}{A\arabic{figure}}
\setcounter{figure}{0}

\appendix

\section*{Appendices}

\section{Parametric instability rates: theory}
\label{app:theory}

For a BEC in the weakly-interacting superfluid regime, excitations above the BEC mode with momentum $k$ are described in momentum space by the Bogoliubov Hamiltonian, see Eq.~\eqref{eq:H_BdG}: 
\begin{equation}
    H_\text{BdG} =  E_0 + \sum_{q\neq k} \left[\xi(k,q) {+} g \right] a^\dagger_q a_q + g\sum_{q \neq k} (a^\dagger_q a^\dagger_{-q} {+} \mathrm{h.c.}),
    \label{eq:H_BdG_app}
\end{equation}
where $\xi(k,q)=E[k+q]-E[k]$ with $E[k]=-2J\cos(k \dL)$ the lattice dispersion. The dynamics of a Bogoliubov quasiparticle $q$ is therefore governed by the Bogoliubov-de Gennes equations (BdGEs)
\begin{equation}
    i \partial_t \left( \begin{matrix} u_{q} \\ v_{q} \end{matrix} \right)=
\left(
\begin{array}{cc}
\xi({k, q}) + g  & g \\ -g & -[\xi({k, -q}) + g]  
\end{array}
\right)
\left(
\begin{array}{c}
u_{q} \\ v_{q}
\end{array}
\right).
\label{eq:HBogo}
\end{equation}
Diagonalizing the BdG matrix in Eq.~\eqref{eq:HBogo} provides the phonon modes and their dispersion $E_\text{ph}(k,q)$.

Exposing this optical lattice to a periodic external force $F(t)$ that couples to the density amounts, in the Peierls gauge, to adding a periodic time dependence in the momentum, $k{\to} k(t)=k_0-f(t)$ with $f(t)=\int^t F(\tau)d\tau$. The time-averaged dispersion $\overline{E[k (t)]}$ [see Eq.~\eqref{eq:AvEnergy_Medium}] describes, at leading order, how the drive renormalizes the lattice dispersion via an effective tunneling $\Jeff$. 
The dynamics of Bogoliubov quasiparticle excitations on top of the driven BEC is described by Eq.~\eqref{eq:HBogo} with substitution $\xi(k, q)\rightarrow\xi[k(t), q]=E[k(t)+q] -E[k(t)]$.\\

A quantitative method to capture parametric instabilities in the system was developed in~\cite{lellouch2017}. In brief, by introducing a change of basis which diagonalizes the time-averaged part of Eq.~\eqref{eq:HBogo}, the BdGEs can be recast into the following form 
\begin{equation}
i \partial_t \left( \begin{matrix} \tilde{u}'_{q}  \\ \tilde{v}'_{q} \end{matrix} \right)=\biggl[E_\text{ph}^{av}\hat{\mathbf{1}}+\hat{W}_\mathbf{q}(t)  +\dfrac{g}{E_\text{ph}^{av}}\left( \begin{matrix} 0 & \tilde{h}_q(t)\mathrm e^{-2iE_\text{ph}^{av}t}        \\ -\tilde{h}_q(t)\mathrm e^{2iE_\text{ph}^{av}t} & 0 \end{matrix} \right)\biggr] \left( \begin{matrix} \tilde{u}'_\mathbf{q}  \\ \tilde{v}'_\mathbf{q} \end{matrix} \right) 
\label{eq:BdGEeffmod}
\end{equation}
where we have used the shortcut notation $E_\text{ph}^{av}$ to denote $E_\text{ph}^{av}(k_0,K,q)$, which is the Bogoliubov dispersion associated with the time-averaged Hamiltonian, within the Bogoliubov approximation,
\begin{eqnarray}
   E_\text{ph}^{av}(k_0 ,q ; K) & & =  \sqrt{\overline{\xi[k(t),q]}(\overline{\xi[k(t),q]}+2g)}
\end{eqnarray}
In the specific case where the condensate is initially prepared in the effective ground state of the driven lattice, i.e. $k_0\dL=\arg(\Jeff)$, which is the situation considered in this work, we find that $ E_\text{ph}^{av}$ is a real-valued, positive quantity given by
\begin{eqnarray}
   E_\text{ph}^{av}(K,q) & & =  \sqrt{4|\Jeff|\sin^2(q\dL/2)[4|\Jeff|\sin^2(q\dL/2)+2g]}\nonumber
\end{eqnarray}
$W_{q}(t)$ is a diagonal matrix of zero average over one driving period which plays no role in the following; and the real-valued function $\tilde{h}_q(t)$ is the time-varying part of the function
\begin{equation}
    h_{q}(t)  =  \frac{1}{2}\left(\xi[k(t),q] + \xi[k(t),-q]\right) \equiv \sum_n h_{q,n}\; e^{-i n\omgd t},
    \label{eq:hq-general}
\end{equation}
i.e. $\tilde{h}_q(t)=\sum_{n\neq 0} h_{q,n}\; e^{-i n\omgd t}$. Denoting $c_n(q)$ the Fourier coefficients of the function $\xi[{k}(t), {q}]$, we can rewrite $\tilde{h}_q(t)=\frac{1}{2}\sum_{n\neq 0}\left[c_n({q})+c_n(-{q}) \right]e^{-in\omgd t}$.

As shown in~\cite{lellouch2017,book_LL}, Eq.~\eqref{eq:BdGEeffmod} takes the form of an effective parametric oscillator model, displaying parametric instability as soon as one harmonic $n\omgd$ of the modulation $h_\mathbf{q}(t)$ is resonant with one of the energies $E_\text{ph}^{av}(k_0 ,q ; K)$ ($n\omgd\approx2E_\text{ph}^{av}(k_0 ,q ; K)$). The instability rate $s_{q,n}$ associated with such parametric resonance for a given quasiparticle momentum $q$ and Fourier mode $n$ can be computed from a standard perturbative treatment of the parametric oscillator~\cite{book_LL,lellouch2017}; in particular, both the magnitude of the instability and the width of the resonance are shown to scale proportionally with the strength of the coupling, $|h_{q,n}|$.
The total instability rate of the system is given by $\Gamma=\max_{{q},n}s_{q,n}$.\\

\subsection{Pulsed drive: perfect annihilation of instabilities 
\label{sec:A_Pulsed_Theory}}

A sufficient condition to suppress any parametric instability in the system is to select a periodic drive such that $h_{q}(t)\equiv 0$ for all quasimomenta $q$, which is the case if $c_n(-{q})=-c_n({q})$ for all $n\neq 0$ and all ${q}$. 
This precisely happens for the pulsed drive considered in Eq.~\eqref{eq:Pulsed_Micromotion} in the limit $M\to\infty$. In this case, the micromotion is given by $k(t) = k_0 + (K/\dL)\text{sign}[\sin(\omgd t)]$ so a straightforward calculation gives $c_{n\neq 0}(q) = (i n\pi)^{-1}((-1)^n -1)\sin K\sin (q\dL)$. Because this is an odd function of $q$, it ensures the fulfillment of the condition $h_{q}(t)\equiv 0$. Therefore, a weakly interacting bosonic system subject to the linear pulsed drive above is free from parametric instabilities.

We note that a weaker condition to suppress the instability rate may be formulated, by demanding that $c_{n^\ast}(-{q})=-c_{n^\ast}({q})$ only for the resonant mode(s) $n^\ast$. In such cases, however, one has to check the behavior of two- and higher-order photon absorption processes. In addition, the effects of the resonance width need to be analyzed as well~\cite{lellouch2017}. In fact, one can go one step further, and demand that the instability is absent only for the maximally-unstable mode $q^\ast$ which dominates the instability rate: $c_{n^\ast}(-{q}^\ast)=-c_{n^\ast}({q}^\ast)$. These considerations can become useful in the engineering of designed drives with suppressed instability rates, e.g., by fine-tuning their Fourier coefficients; alternatively, one can explore a wide range of parameter regimes to identify sweet spots of vanishingly small instability rates, as we do in the following.

\subsection{Two-tone drives with tunable amplitudes and phases: restoring stability in strongly-normalized regimes \label{sec:A_Twotone_Theory}}

We consider here the case of a single-band optical lattice driven by a two-tone drive, as defined in Eq.~\eqref{eq:micromotion2}
\begin{align}\label{eq:micromotion2_app}
\hbar k(t) = \hbar k_0  + \frac{\hbar\kL}{\pi} \left[ K_1 \sin(\omega_1 t) + K_2 \sin(\omega_2 t + \varphi) \right]
\end{align}
where $\varphi$ is the relative phase between the two components, and $\omega_2$ is chosen to be an integer multiple of $\omega_1$, i.e. $\omega_2=N\omega_1$ with $N\in \mathbb{N}^*$. The experimental situation considered in this work corresponds to the case $N=2$. 

In the absence of interactions, the effective Floquet Hamiltonian can be obtained at lowest order by taking the time-average of $E[k(t)]=-2J\cos[k(t)\dL]$. This produces an effective dispersion analogous to Eq.~\eqref{eq:AvEnergy_Medium}, but with an renormalized tunneling
\begin{equation}
\Jeff(K_1,K_2,\varphi,N)=J \sum_{p=-\infty}^{\infty} \mathcal{J}_{-Np}(K_1)\mathcal{J}_{p}(K_2)\e^{ip\varphi}
\label{eq:Jeff_2freq_app}
\end{equation}
with $\mathcal{J}_l(z)$ the $l$-th Bessel function of the first kind. Higher-order corrections to the effective Hamiltonian would generally give rise to long-range hopping. Eq.~(\ref{eq:Jeff_2freq_app}) is real-valued in the case $\varphi=0$, in which case the ground state occurs at $k_0=0$ for parameters such that $\Jeff(K_1,K_2,\varphi,N)>0$ and $k_0=\kL$ for parameters such that $\Jeff(K_1,K_2,\varphi,N)<0$. However, for an arbitrary phase $\varphi$ between the two modulations, the effective tunneling is generally complex and condensation occurs at $k_0$ with $k_0 \dL=\mathrm{arg}[\Jeff(K_1,K_2,\varphi,N)]$.\\

\begin{widetext}
To evaluate the instabilities in the system, we follow the general approach above, which allows to cast the BdGEs under the form of Eq.~\eqref{eq:BdGEeffmod}, where the function $\tilde{h}_q(t)$ is now given by $\tilde{h}_q(t)=4J\sin^2(q\dL/2)\sum_{n\neq 0} A_n(K_1,K_2,\varphi,N)  \mathrm e^{in\omgd t},$
where
\begin{equation}
A_n(K_1,K_2,\varphi,N)=\frac{1}{2}\sum_{p=-\infty}^\infty \left( \mathcal{J}_{n-Np}(K_1)e^{i(p\varphi-\mathrm{arg}[\Jeff(K_1,K_2,\varphi,N)])}+\mathcal{J}_{-n-Np}(K_1)e^{-i(p\varphi-\mathrm{arg}[\Jeff(K_1,K_2,\varphi,N)])} \right) \mathcal{J}_{p}(K_2).
\end{equation}
To obtain analytical estimates of instability rates, we further treat each Fourier mode $n$ of the drive independently, which allows to recast the dynamics of a given quasiparticle momentum $q$ subjected to a given mode $n$ into the form of a parametric oscillator model
\begin{align}
&i \partial_t \! \left( \begin{matrix} \tilde{u}'_q  \\ \tilde{v}'_q \end{matrix} \right)=\biggl[ E_\text{ph}^{av}\hat{\mathbf{1}}
+\hat{W}_q(t) 
+\!\dfrac{\alpha_{q,n}E_\text{ph}^{av}}{2}\left( \begin{matrix} 0 & \cos(n\omgd t+\theta_n)\mathrm e^{-2iE_\text{ph}^{av}t}        \\ -\cos(n\omgd t+\theta_n)\mathrm e^{2iE_\text{ph}^{av}t} & 0 \end{matrix} \right)\biggr]\! \left( \begin{matrix} \tilde{u}'_q  \\ \tilde{v}'_q \end{matrix} \right),
\label{eq:BdGEO2}
\end{align}
where we have used again the shortcut notation $E_\text{ph}^{av}$ to denote the Bogoliubov dispersion associated with the time-averaged Hamiltonian, $E_\text{ph}^{av}=E_\text{ph}^{av}(k_0,q;K_1,K_2,\varphi,N)$, and where 
\begin{equation}
\alpha_{q,n}=16J \sin^2(q\dL/2) \frac{g}{(E_\text{ph}^{av})^2}\Big| A_n(K_1,K_2,\varphi,N)\Big| 
\label{eq:alphaqm}
\end{equation}
represents the coupling strength of the effective parametric oscillator and $\theta_n=\arg\Big[A_n(K_1,K_2,\varphi,N)\Big]$
is an additional phase that appears only in the case where $\varphi\neq 0$.
\end{widetext}

The parametric oscillator model Eq.~\eqref{eq:BdGEO2} can be solved perturbatively in $\alpha_{q,n}$~\cite{lellouch2017,book_LL}. At first order, the instability rate is given by
$$s_{q,n}=\alpha_{q,n}E_\text{ph}^{av}/{4}\sqrt{1-2(n\omgd-2E_\text{ph}^{av})/\alpha_{q,n}E_\text{ph}^{av}}.$$ 
It is non-zero as soon as the drive frequency is close to the resonance $n\omgd\approx2 E_\text{ph}^{av}$. Corrections beyond this first order result can be obtained perturbatively~\cite{book_LL,lellouch2017}. As discussed in~\cite{lellouch2017}, this perturbative approach however breaks done in regimes where the effective tunneling vanishes and the Bogoliubov band $E_\text{ph}^{av}$ becomes flat. An accurate solution can nevertheless be found numerically by numerically solving the BdGEs, as presented on Fig.~\ref{Fig:TwoFreq_PhononGrowth}.

Through $\alpha_{q,n}$ and $E_\text{ph}^{av}$, the instability rate exhibits a non-trivial dependence of the drive tuning parameters $K_1$, $K_2$, $\varphi$ and $N$, which differs from the dependence of the effective tunneling Eq.~\eqref{eq:Jeff_2freq_app} on these parameters; exploring the parameter space allows to find sweet spots where a given effective tunneling can be realized at significantly suppressed instability rates.

\subsection{Reducing $\Gamma$ with two-tone drive \label{sec:TwoFreq_Motivation}}

\begin{figure} [b]
\includegraphics[width=\columnwidth]{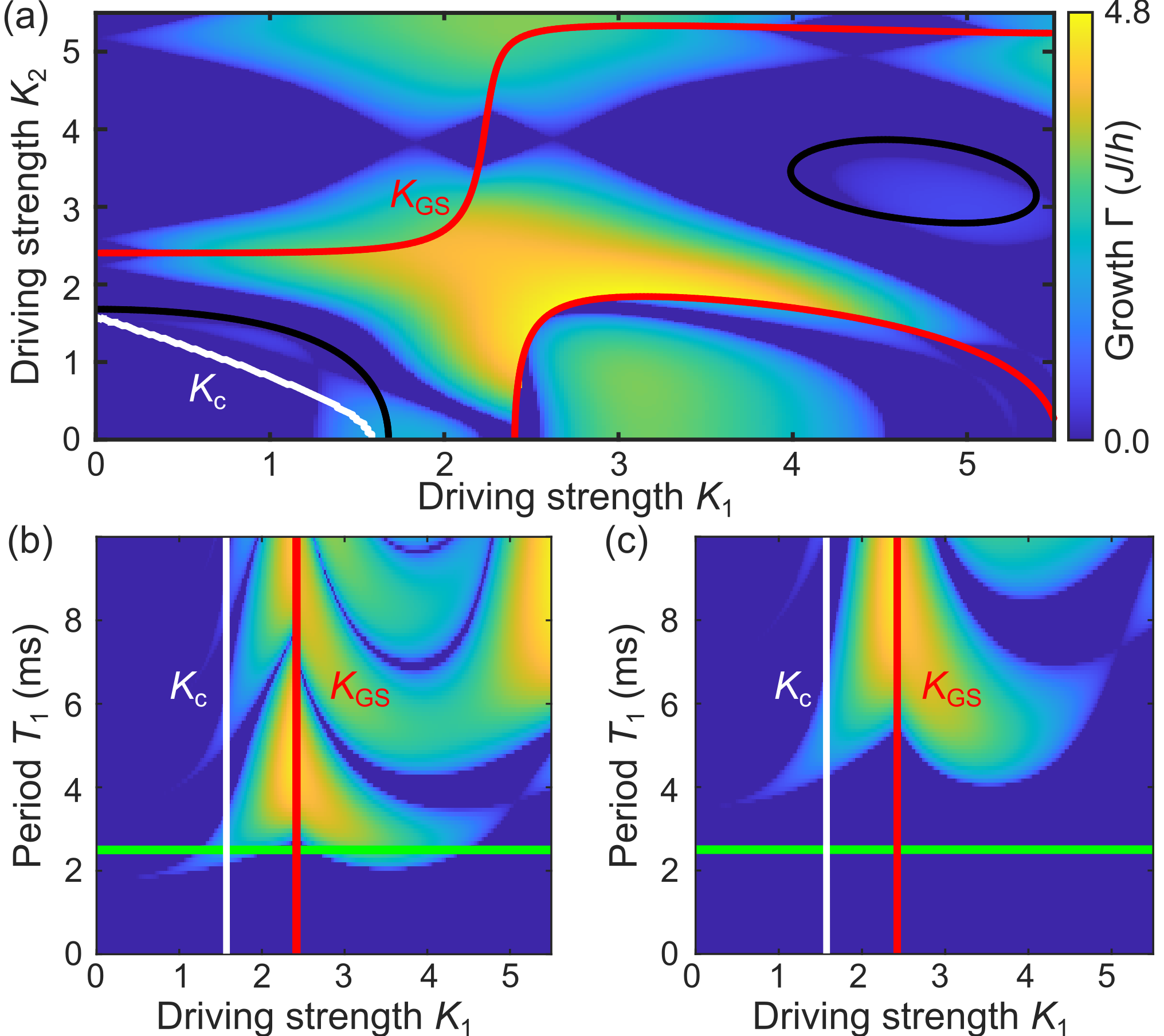}
\caption{Calculated phonon growth rate. (a) Phonon growth rate $\Gamma$ for mode $q=\kL$ and driving period $T_1=2.5\,$ms with parameters $V=8\,\Er$, $g/J=15$. Red lines indicate the change of ground state at $K_\text{GS}$ and the white line marks the critical driving strengths $K_c$ beyond which negative-mass instabilities occur. The black lines indicate $|\Jeff/J|=0.4$. (b,c) Dependence of the phonon growth rate along the axes in (a) on the driving period $T_1$, i.e. for ($K_1$,$K_2=0$) in (b) and ($K_1=0$,$K_2$) in (c). The driving period for (a) is indicated by green lines. \label{Fig:TwoFreq_CalcGrowth}} 
\end{figure}

The ability to tune the phonon growth rate with a two-tone drive in Sec.\,\ref{sec:TwoFreq_Drive} arises from the detailed structure of $\Gamma$ in the $(K_1, K_2)$-plane. This structure, calculated in Fig.\,\ref{Fig:TwoFreq_CalcGrowth}(a) with the BdG equations, can be understood by examining $\Gamma$ along the two axes where only one driving frequency is present. Along the $K_1$-axis, the growth rate exhibits a characteristic time dependence with a sequence of higher-order resonances and a strong increase for $K_1>K_c$ due to negative-mass instabilities~\cite{creffield2009, lellouch2018, dicarli2023, cruickshank2024} [Fig.\,\ref{Fig:TwoFreq_CalcGrowth}(b)]. A similar time dependence appears along the $K_2$-axis, but scaled in time by a factor of two due to the doubled driving frequency [Fig.\,\ref{Fig:TwoFreq_CalcGrowth}(c)]. Green lines indicate the driving period used in Fig.\,\ref{Fig:TwoFreq_CalcGrowth}(a).

The two-dimensional map in Fig.\,\ref{Fig:TwoFreq_CalcGrowth}(a) can be viewed as an interpolation between these limiting cases, with intermediate values for the combined micromotion. The points $K_c$ and $K_\text{GS}$ turn into curves across the $(K_1, K_2)$-plane. Regions with enhanced growth arise where the fundamental and high-order resonances intersect this plane, which is set by the driving period. This structure enables tuning of $\Gamma$ while keeping the desired value of $\Jeff$ by selecting trajectories with reduced phonon growth, e.g., along the black line in Fig.\,\ref{Fig:TwoFreq_CalcGrowth}(a).

\subsection{Reducing $\Gamma$ with pulsed drive \label{sec:Pulsed_Motivation}}

\begin{figure} 
\includegraphics[width=\columnwidth]{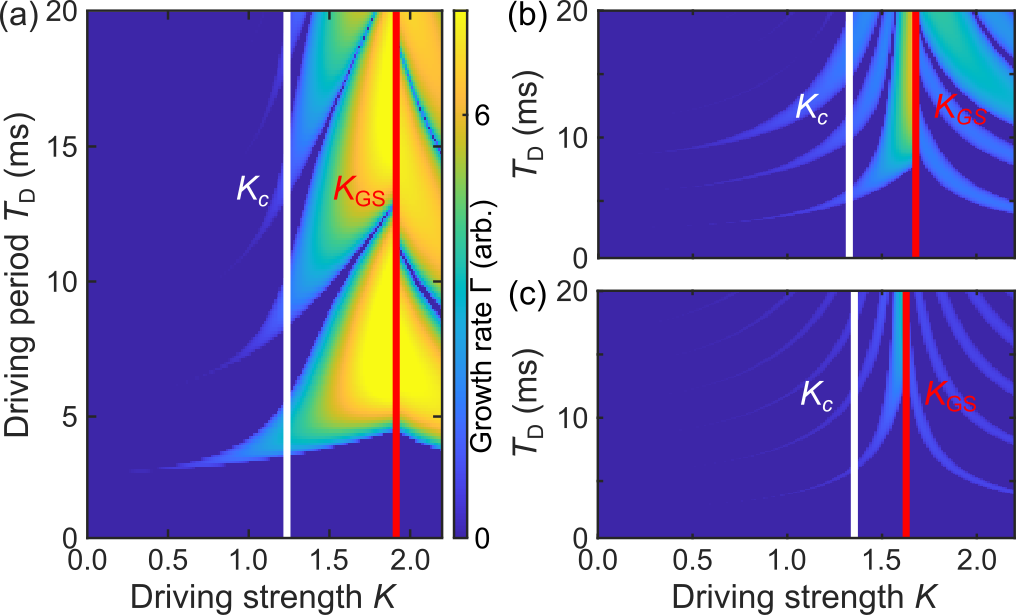}
\caption{Growth of phonon modes for pulsed drive. (a) Calculated growth rate $\Gamma$ using the BdG equations for the fundamental mode $M=1$ with $V_0=10\,E_\text{r}$, $g=15 J$, $q=\kL$, with $K$ defined in Eq.\,\eqref{eq:Pulsed_Micromotion}. The white line indicates the critical driving strength $K_c$ beyond which negative-mass instabilities occur, while the red line indicates the change of the ground state from $k_0=0$ to $k_0=\kL$. The growth rate decreases drastically with increasing number of harmonics with $M=5$ (b) and $M=13$ (c). Same color map for all panels.
\label{Fig:GrowthRate_Sine_A}}
\end{figure}

The suppression of phonon growth under pulsed driving arises from two mechanisms: the elimination of parametric instabilities, due to the vanishing of the resonant Fourier coefficients $h_{q,n}$ as discussed in  Sec.\,\ref{sec:Exp_PhononModes} and Appendix~\ref{sec:A_Pulsed_Theory}, and the disappearance of negative-mass instabilities, because the micromotion no longer crosses into regions of negative effective mass [Sec.\,\ref{sec:Pulsed_Phonons}]. Figure~\ref{Fig:GrowthRate_Sine_A} illustrates the resulting reduction of the calculated growth rate as the number of harmonics $M$ increases and the driving function approaches the pulsed-driving limit.

For reference, we provide $\Gamma$ for the case of sinusoidal driving $M=1$ [Fig.\,\ref{Fig:GrowthRate_Sine_A}(a)]. As before, red and white lines indicate the change of ground state at $K_\text{GS}$ and the critical driving strengths $K_c$, respectively. As the number of harmonics increases, $\Gamma$ decreases drastically, as illustrated for $M=5$ and $13$ in Fig.\,\ref{Fig:GrowthRate_Sine_A}(b,c). Furthermore, $K_c$ and $K_\text{GS}$ converge toward one another, approaching $K=\pi/2$ in the limit of large $M$. Consequently, the ground state switches from $k_0=0$ to $\kL$ at $K_c$, occurring precisely when the micromotion would otherwise enter the negative-effective-mass region of the Brillouin zone.

\section{Experimental methods\label{sec:Details_Setup}}

\subsection{Measuring the micromotion\label{sec:Reference_Frames}}

A direct measurement of the atoms' quasi-momentum distribution using band-mapping techniques is challenging for our driving scheme. The quasi-momentum is defined in the lattice frame, whereas our absorption images are recorded in the laboratory frame, in which the lattice itself is oscillating due to the drive. Instead, we studied the atoms' micromotion by measuring their center-of-mass momentum, $\hbar \pcm$. In the lattice frame, the momentum distribution $n(p_\text{lat}, t)$ can be extracted from the interference pattern after expansion~\cite{arimondo2012a}
\begin{align*}\label{eq:momentum_latticeframe}
n(p_\text{lat}, t) &= |\widetilde{w}(p_\text{lat})|^2 \\
& \times \sum_{r, s} \langle  \psi | \hat{b}^\dagger_r \hat{b}_s | \psi \rangle  \exp\left[ i (r - s) (p_\text{lat}\dL+ k[t] \dL) \right],
\end{align*}
where $\hat{b}_s$ denotes the annihilation operator for an atom at lattice site $s$. The sum captures the contribution of atoms at individual lattice sites ($r,s$), and the Fourier transform $\widetilde{w}(p_\text{lat})$ of a Wannier function provides a broad momentum-space envelope. The resulting momentum distribution shows peaks spaced by $2\hbar \kL$, whose relative weights follow the envelope $\widetilde{w}(p_\text{lat})$. The positions of these peaks shift dynamically in time, directly reflecting the atoms' micromotion $k[t]$.

In the laboratory frame, the momentum distribution changes due to the relative motion between the lab and the oscillating lattice. Since the lattice velocity is always opposite to the micromotion of the atoms, the lab-frame distribution becomes~\cite{arimondo2012a}
\begin{align*}
n_{\text{lab}}(p_{\text{lab}}, t) &= \left| \widetilde{w}\left(p_{\text{lab}} -k[t] \right) \right|^2 \\
 &\quad \times \sum_{r,s}  \langle \psi | \hat{b}^\dagger_r \hat{b}_s | \psi \rangle \exp\left[ i(r - s) p_{\text{lab}} \dL \right].
\end{align*}
In this frame, the interference pattern retains fixed peak positions corresponding to a momentum $\hbar k_0$, while $\widetilde{w}$ oscillates in time. This temporal modulation of the envelope can be directly tracked by monitoring the center-of-mass momentum $\hbar\pcm(t)$ which oscillates with the center of $\widetilde{w}$ given by $k[t]$. Note that additional non-inertial forces, e.g., due to gravity or magnetic field gradients, can lead to a shift of the interference peaks.

\begin{figure}
\includegraphics[width=\columnwidth]{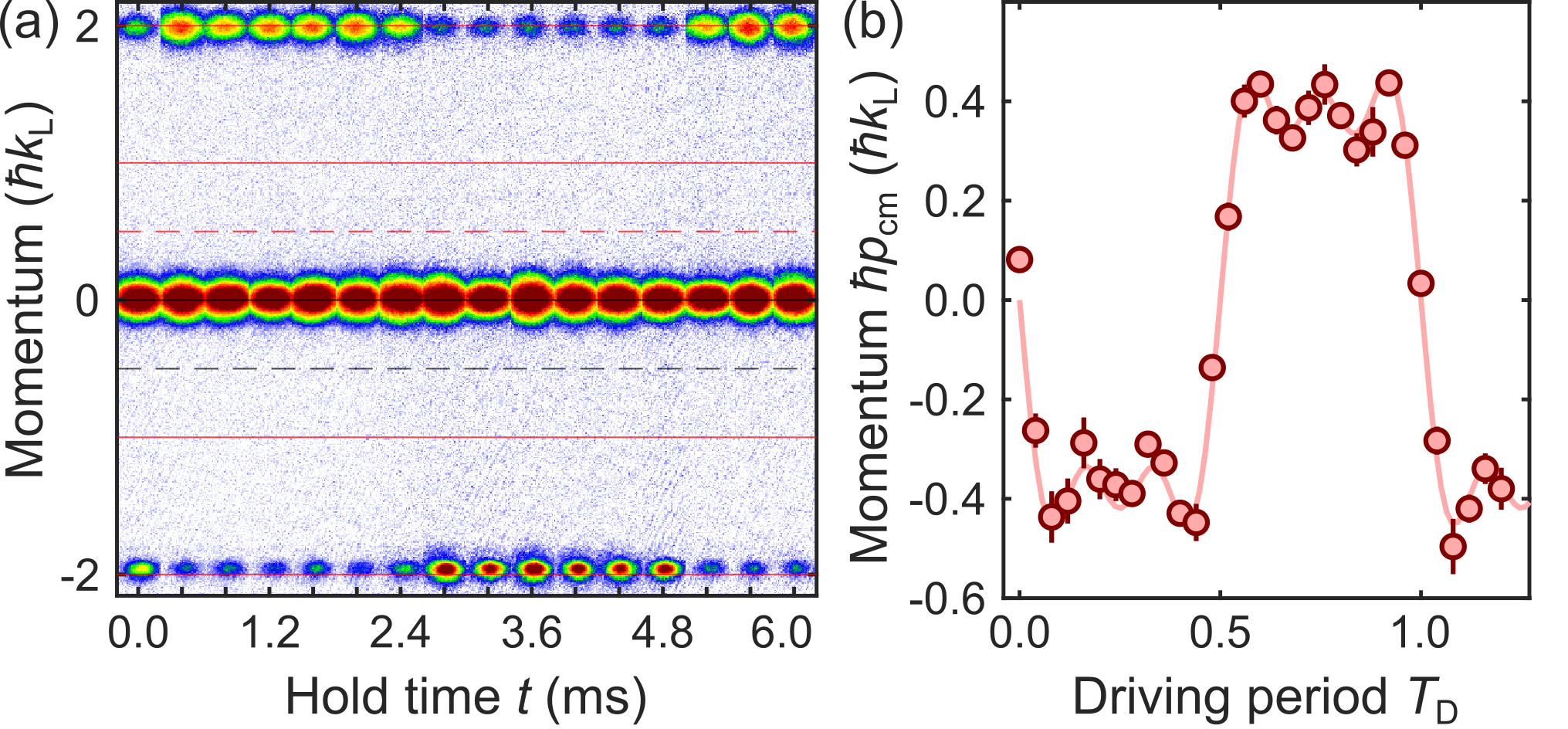}
\caption{Micromotion for pulsed driving. (a) Absorption images of the density profile in momentum space after increasing the driving duration $t$, parameters $M=5$, $V_0=7\,E_\text{r}$, $a_s=2\,a_0$, $K=0.9$. (b) The measured center-of-mass momentum $\hbar \pcm$ (red circles) of the cloud in (a) shows good agreement with the micromotion $k(t)$ calculated with Eq.\,\eqref{eq:Pulsed_Micromotion} (red line).\label{Fig:PulsedMicromotion_A}}
\end{figure}

We used this method to validate our experimental implementations of the micromotion for both pulsed and two-tone driving schemes. As an example, Fig.\,\ref{Fig:PulsedMicromotion_A}(a) shows absorption images of the atomic momentum distribution taken after increasing durations $t$ of a pulsed drive. The modulation of the envelope is directly observable through the varying intensity of the interference peaks, while their positions remain fixed. The extracted center-of-mass momentum $\hbar\pcm$ agrees well with the expected micromotion for $M=5$, as calculated using Eq.\,(\ref{eq:Pulsed_Micromotion}) [Fig.\,\ref{Fig:PulsedMicromotion_A}(b)].

\subsection{Measuring $\Jeff$ for pulsed driving\label{sec:Pulsed_MeasureJeff_A}}

\begin{figure} 
\includegraphics[width=\columnwidth]{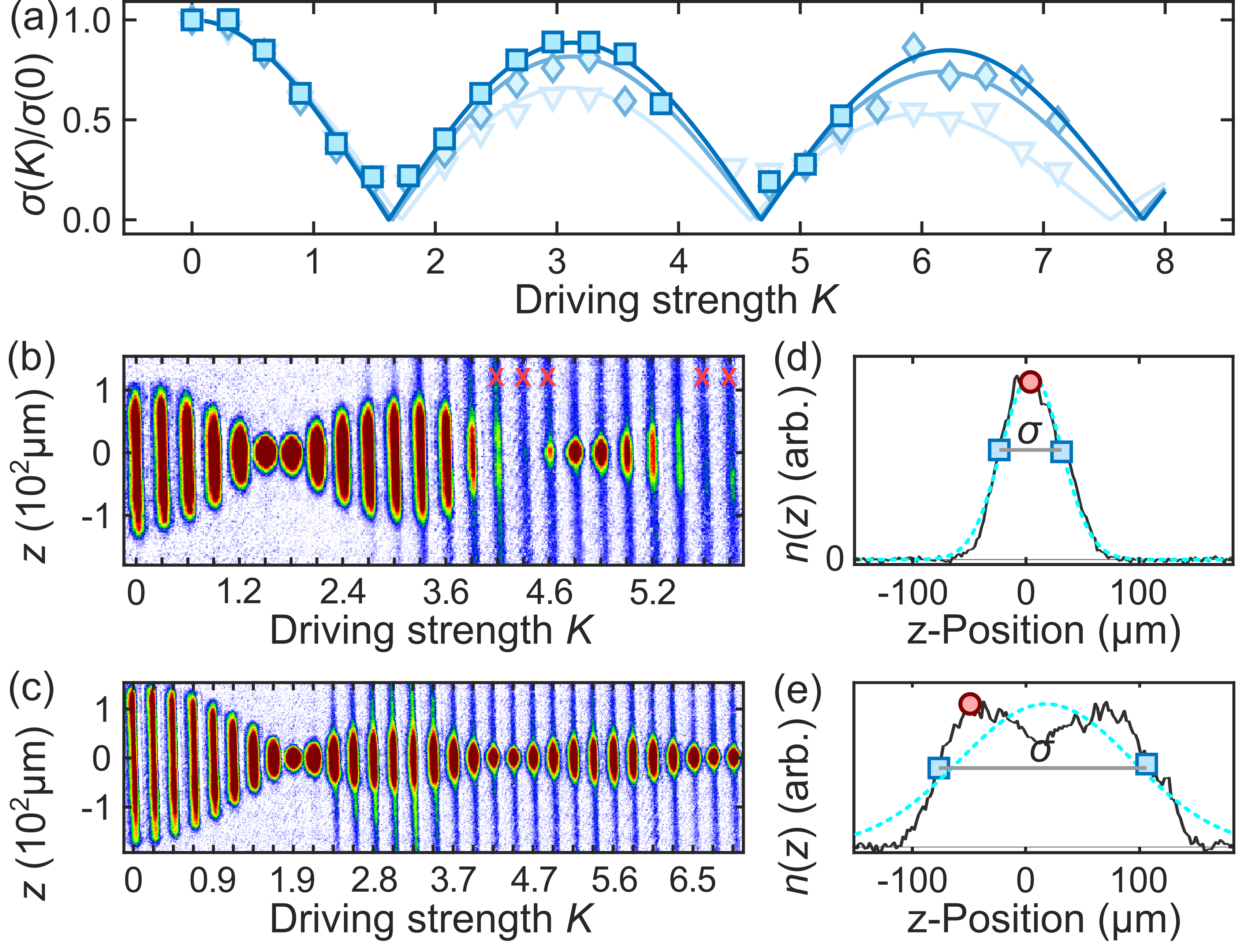}
\caption{Tunneling rate under pulsed driving for different numbers of harmonics. (a) Normalised expansion width after $700$\,ms for thirteen (squares), seven (diamonds), and three (triangles) harmonics; other parameters as in Fig.\,\ref{Fig:SquareDrive_Jeff}(a). Solid lines show the calculated values of $|\Jeff/J|$.
(b) Averaged \emph{in situ} absorption images of the cloud width for $M=11$ harmonics. Red crosses mark data sets affected by atom loss due to band excitations. (c) Averaged \emph{in situ} absorption images for sinusoidal driving. (d) Vertically integrated 1D density profile of a non-interacting cloud after $600\,$ms of sinusoidal driving with $K=2.1$, $\TD=6\,$ms, $V_0=11\,\Er$. (e) Density profile of a spreading wave packet after $800\,$ms. Red circles indicate the maximum density $n_\text{max}$, blue squares the points where $n=0.6\,n_\text{max}$, and blue lines Gaussian fits. The initial cloud width at $t=0$ is $15.6\,\upmu$m.}\label{Fig:Pulsed_MeasureJeff_A}
\end{figure}

For pulsed driving, the effective tunneling rate $\Jeff(K)$ depends on the number of harmonics $M$ in the drive's waveform. In the limiting cases, $\Jeff(K)$ changes from $J\mathcal{J}_0\left(4K/\pi\right)$ for a single harmonic ($M=1$) to $J\cos(K)$ in the limit of many harmonics ($M\rightarrow\infty$). For intermediate values of $M$, we calculated $\Jeff(K)$ numerically using the following procedure: (i) We first determined the micromotion $k(t)$ for the given driving parameters using Eq.\,(\ref{eq:Pulsed_Micromotion}). (ii) We then calculated the time-averaged energy $\bar{E}(k_0)=\overline{E[k(t),q]}$ over one full driving cycle and (iii) finally determined $\Jeff$ from the bandwidth of the effective dispersion relation which is $4\Jeff$.

To measure $\Jeff(K)$, we analyzed the wave packet's spreading as a function of the driving strength $K$. The lattice depth was increased from $8\,\Er$ to $11\,\Er$ for larger number of harmonics ($M>11$) to reducing band excitations at large values of $K$. We avoided self-trapping and other many-body effects by tuning the $s$-wave scattering length to $a_\text{s}=+2(1)\,a_0$. Tunneling along the lattice direction was initialized by removing the trapping beam $D_1$ in $3\,$ms and starting the shaking of the lattice potential. The measurements in Fig.\,\ref{Fig:Pulsed_MeasureJeff_A} use a driving period of $\TD=5\,$ms and a driving duration of $700\,$ms. Absorption images were taken almost in-situ, with a short free expansion time of $2\,$ms.

The normalized tunneling rate $|\Jeff(K)/J|$ matches well to the relative cloud width $\sigma(K)/\sigma(0)$ for long driving durations~\cite{lignier2007}. We employed two methods to extract $\sigma$ from the density profiles shown in Fig.\,\ref{Fig:Pulsed_MeasureJeff_A}. In the case of a strong suppression of tunneling, the density profiles remained approximately Gaussian, allowing us to extract $\sigma$ using a Gaussian fit [cyan line in Fig.\,\ref{Fig:Pulsed_MeasureJeff_A}(d)]. For stronger tunneling, however, the cloud exhibited significant spreading, and the density profile developed a bimodal shape, rendering Gaussian fits unreliable [Fig.\,\ref{Fig:Pulsed_MeasureJeff_A}(e)]. While previous studies have addressed this by fitting a sum of two Gaussians~\cite{lignier2007}, we instead determined $\sigma$ at a fixed fraction of the peak density $n_\text{max}$ [gray horizontal lines in Fig.\,\ref{Fig:Pulsed_MeasureJeff_A}(d) and (e)]. For pulsed driving, we used $60\%$ of $n_\text{max}$ [blue data in Fig.\,\ref{Fig:SquareDrive_Jeff}(a)], whereas for sinusoidal driving, the width was determined at $40\%$ [red data in Fig.\,\ref{Fig:SquareDrive_Jeff}(a)]. We verified that the choice of $n/n_\text{max}$ does not influence the measured value of $|\Jeff(K)/J|$.

We observed a strong atom loss for certain driving parameters, indicated by red crosses in Fig.\,\ref{Fig:Pulsed_MeasureJeff_A}(b). This loss arises from band excitations induced by strong acceleration during the drive, which leads to Landau-Zener transitions. For pulsed driving, the peak acceleration increases with $K$, $\omgd$, and $M$, whereas the critical acceleration for Landau-Zener transitions is set by the lattice depth $V_0$. These band excitations occur for specific parameter combinations, leading to a drop in atom number over the long driving duration. We excluded these parameters from our measurement of $\Jeff$.

\vfill

\end{document}